\documentclass[preprint2]{aastex}
\pdfoutput=1
 
\usepackage{caption} 
\usepackage{natbib}
\usepackage{graphicx}
\usepackage{amsmath}
\usepackage{txfonts}
\usepackage{amssymb}
\usepackage{subfig}
\usepackage{rotating}

\shorttitle{The Host Galaxies and Narrow Line Regions of Four Double-Peaked $\left[ \textrm{OIII} \right]$ AGN}
\shortauthors{Villforth \& Hamann}
\begin{document}

\title{The Host Galaxies and Narrow Line Regions of Four Double-Peaked $\left[ \textrm{OIII} \right]$ AGN}

\author{Carolin Villforth$^{1,2}$ \& Fred Hamann $^{1}$}
\affil{$^{1}$ Department of Astronomy, University of Florida, 32611 Gainesville, Florida, United States}
\affil{$^{2}$ Scottish Universities Physics Alliance (SUPA), University of St Andrews, School of Physics and Astronomy, North Haugh, KY16 9SS, St. Andrews, Fife, UK}

\keywords{galaxies: active -- galaxies: interactions -- galaxies: kinematics and dynamics -- quasars: general}

\begin{abstract}
Major gas-rich mergers of galaxies are expected to play an important role in triggering and fuelling luminous AGN. The mechanism of AGN fuelling during mergers however remains poorly understood. We present deep multi-band ($u/r/z$) imaging and long slit spectroscopy of four double-peaked $\left[ \textrm{OIII} \right]$ emitting AGN. This class of object is likely associated with either kpc-separated binary AGN or final stage major mergers, though AGN with complex narrow-line regions are known contaminants. Such objects are of interest since they represent the onset of AGN activity during the merger process. Three of the objects studied have been confirmed as major mergers using near-infrared imaging, one is a confirmed X-ray binary AGN. All AGN are luminous, radio-quiet to radio-intermediate and have redshifts of 0.1 $<$ z $<$ 0.4. Deep $r$-band images show that a majority (3/4) of the sources have disturbed host morphologies and tidal features, while the remaining source is morphologically undisturbed down to low surface brightness limits ($\sim$ 27 mag/arcsec $^{2}$ in $r$). The lack of morphological disturbances in this galaxy despite the fact that is is a close binary AGN suggests that the merger of a binary black hole can take longer than $~$1 Gyr. All AGN hosted by merging galaxies have companions at distances $\leqslant$ 150 kpc. The narrow line regions (NLRs) have large sizes (10 kpc$< r <$100 kpc) and consist of compact clumps with considerable relative velocities between components ($\sim$ 200-650 km s$^{-1}$). We detect broad, predominantly blue, wings with velocities up to $\sim$1500 km s$^{-1}$ in $\left[ \textrm{OIII} \right]$, indicative of powerful outflows. The outflows are compact ($<$5 kpc) and co-spatial with nuclear regions showing considerable reddening, consistent with enhanced star formation. One source shows an offset between gas and stellar kinematics, consistent with either a bipolar flow or a counter-rotating gas disk. In all other sources, the ionized gas generally follows the stars. We are not able to unambiguously identify the sources as binary AGN using our data, X-ray or radio data is required for an unambiguous identification. However, the data still yield interesting results for merger triggering of AGN and time-scales of binary black hole mergers.
\end{abstract}

\section{Introduction}

Supermassive black holes (SMBHs) are now believed to be present in the centres of most if not all massive galaxies \citep[e.g.][and references therein]{kormendy_coevolution_2013}. While most SMBHs do not accrete large amounts of gas, a small fraction of them show strong signs of accretion. These objects are known as Active Galactic Nuclei (AGN).

The processes that trigger and fuel AGN are still poorly understood. Major mergers are often suggested to be triggers of AGN activity \citep[e.g.][]{sanders_ultraluminous_1988,hopkins_cosmological_2008}. While some early studies suggested a link between mergers and AGN \citep{canalizo_quasi-stellar_2001}, other studies had shown much lower incidences of mergers in AGN hosts \citep{bahcall_apparently_1996,bahcall_hubble_1997}. \citet{veilleux_deep_2009} suggested that the connection is between starbursts and disturbed morphologies, rather than AGN and disturbed morphologies. Recent studies using matched control samples found no connection between mergers and AGN \citep{cisternas_secular_2011,boehm_agn_2012,kocevski_candels:_2012,villforth_morphologies_2014}, while some studies do find increased signs of mergers \citep{urrutia_evidence_2008,ramos_almeida_optical_2011}. Most surprisingly, several studies have found increased incidences of AGN in close pairs of galaxies \citep{ellison_galaxy_2008,ellison_galaxy_2013,koss_understanding_2012}. 

Seemingly conflicting results show the importance of understanding the properties of AGN in ongoing mergers. Studying AGN in mergers requires identifying suitable samples. A promising type of object is binary or dual AGN, i.e., those in which two black holes are active. Since binary AGN require the merger of two galaxies massive enough to host powerful AGN, they must be located in mergers of galaxies with similar masses. Binary AGN therefore trace major mergers before the coalescence of the two black holes. Such samples are difficult to identify, especially in powerful AGN that outshine their host galaxies. Besides serendipitous detections \citep{junkkarinen_lbqs_2001,komossa_discovery_2003,koss_chandra_2011}, there have been claimed detections of close sub-pc binary AGN \citep{boroson_candidate_2009,valtonen_massive_2008,valtonen_tidally_2009}, although in many cases the data does not unambiguously support those claims \citep{chornock_sdss_2009,villforth_variability_2010}.

A promising way to identify binary AGN is to select AGN that have narrow line regions (NLRs) with components at two different velocities, typically separated by $\sim 200-500$ km s$^{-1}$ \citep{smith_search_2010,shen_binary_2010}, this method identified sources in which each AGN illuminates separate gas clouds, creating individual NLRs for each AGN. Since sizes for NLRs are typically $\sim$ 100 pc - 1 kpc \citep[e.g.][]{liu_observations_2013,bennert_size_2002,bennert_size_2006,humphrey_integral-field_2010,schmitt_hubble_2003,greene_feedback_2011,fu_extended_2009}, binary AGN with much smaller separations are not identified using this method. However, complex gas kinematics as well as bi-conical in- or outflows can also cause double peaked $\left[ \textrm{OIII} \right]$ emission \citep{fu_mergers_2011,rosario_adaptive_2011,barrows_candidate_2011,fu_nature_2012}. Some double-peaked $\left[ \textrm{OIII} \right]$ AGN have been confirmed as binary AGN in follow-up observations \citep{fu_mergers_2011,fu_nature_2012,fu_kiloparsec-scale_2011,comerford_chandra_2011}, while others are AGN with in- or outflows or kinematically disturbed NLRs \citep{fu_mergers_2011,fu_nature_2012}. However, the properties of the host galaxies of double-peaked $\left[ \textrm{OIII} \right]$ remain poorly understood.

In this paper we present a detailed study of four double-peaked $\left[ \textrm{OIII} \right]$ AGN, identified by \citet{smith_search_2010} and \citet{shen_binary_2010}. Three have been confirmed to be in major mergers by \citet{fu_nature_2012} using near-infrared (NIR) imaging, where they show two distinct cores associated with line emission indicative of an AGN. One of those three has also been confirmed as a binary AGN in the radio \citep{shen_binary_2010}. Additionally, we study one X-ray binary AGN identified by \citet{comerford_chandra_2011}. Deep imaging data in $u,r,z$ as well as medium-resolution long-slit spectroscopy for these four sources were obtained. We will study the host galaxy morphology as well as gas and stellar kinematics across the host galaxies to constrain the merger stages and host galaxy properties. These data will allow us to constrain morphologies, star-formation properties, NLR kinematics, incidence of fast outflows as well as gas and stellar kinematics.

The sample is presented in Section \ref{S:sample}, followed by observations and data reduction in Section \ref{S:obs}. The results for the full sample are presented in Section \ref{S:results}, subdivided into morphology (Section \ref{S:imaging}) and kinematics (Section \ref{S:spec}). The results are discussed in Section \ref{S:discussion}, followed by conclusions in Section \ref{S:conclusions}. Supplemental information is provided in the Appendix. The cosmology used is $H_{0}=70\textrm{km s}^{-1}\textrm{Mpc}^{-1},\Omega_{\Lambda}=0.7, \Omega_{m}=0.3$. Throughout the paper, we use AB magnitudes.

\begin{figure*}
\includegraphics[width=8.5cm]{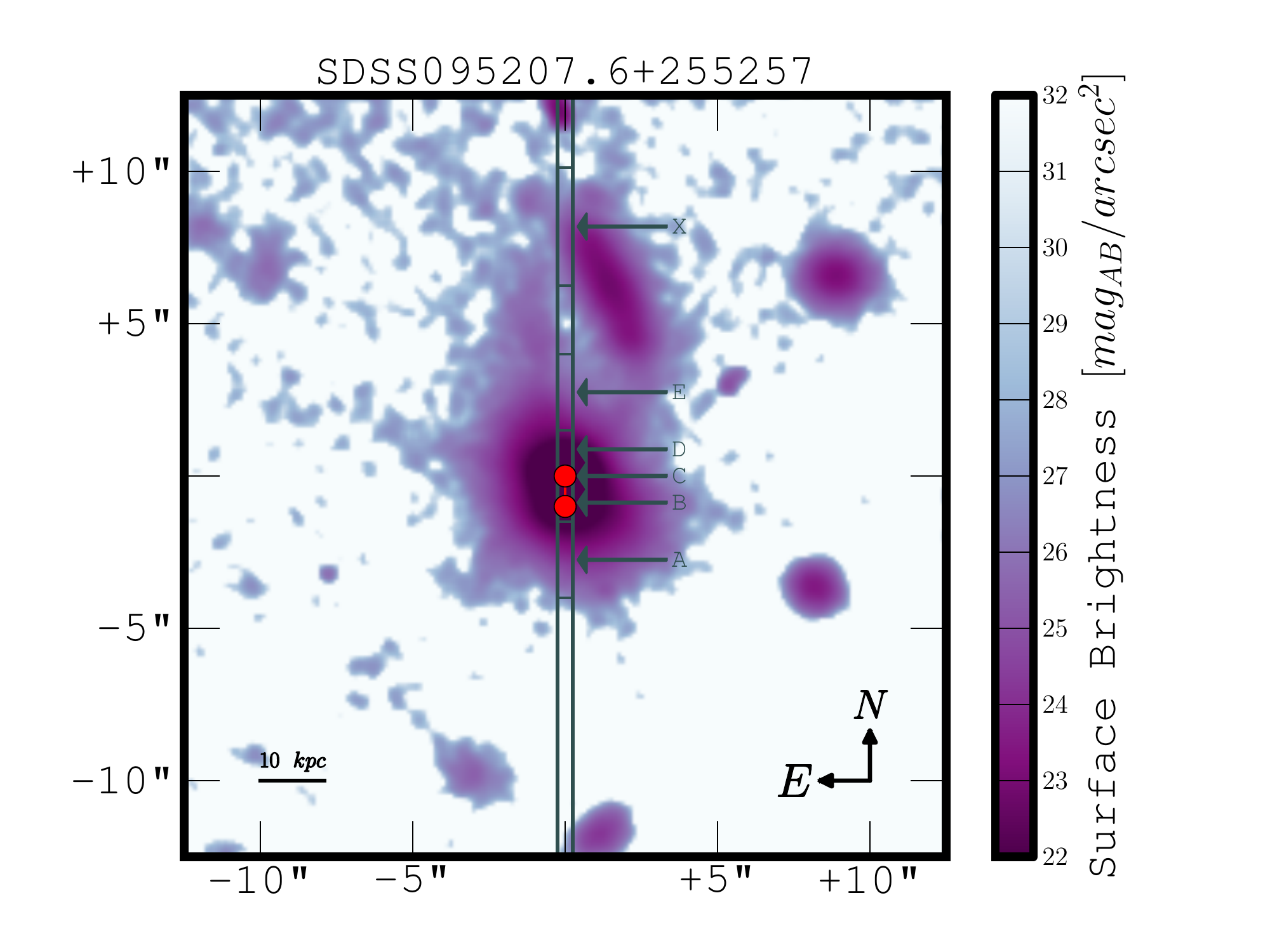}
\includegraphics[width=8.5cm]{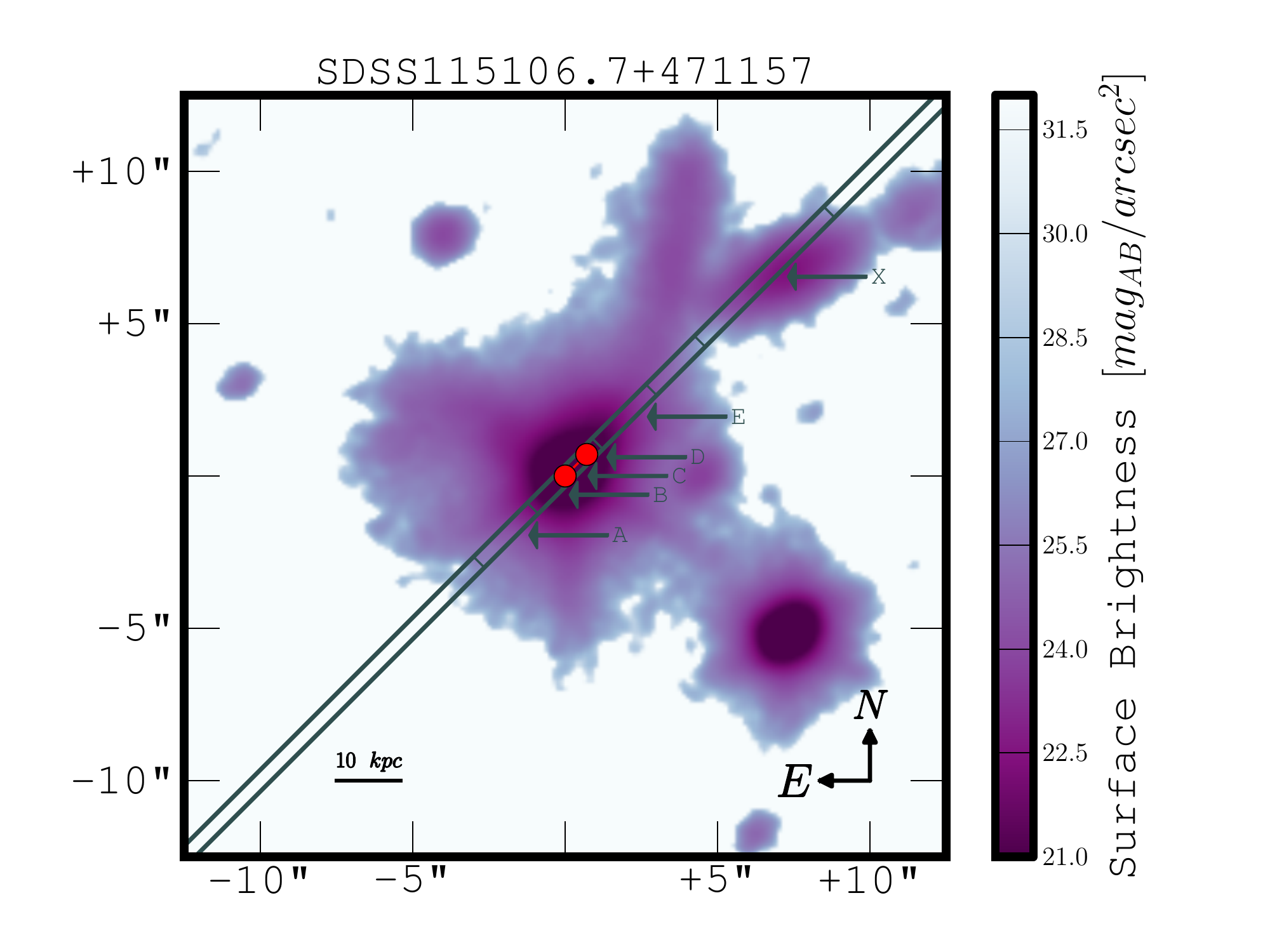}
\includegraphics[width=8.5cm]{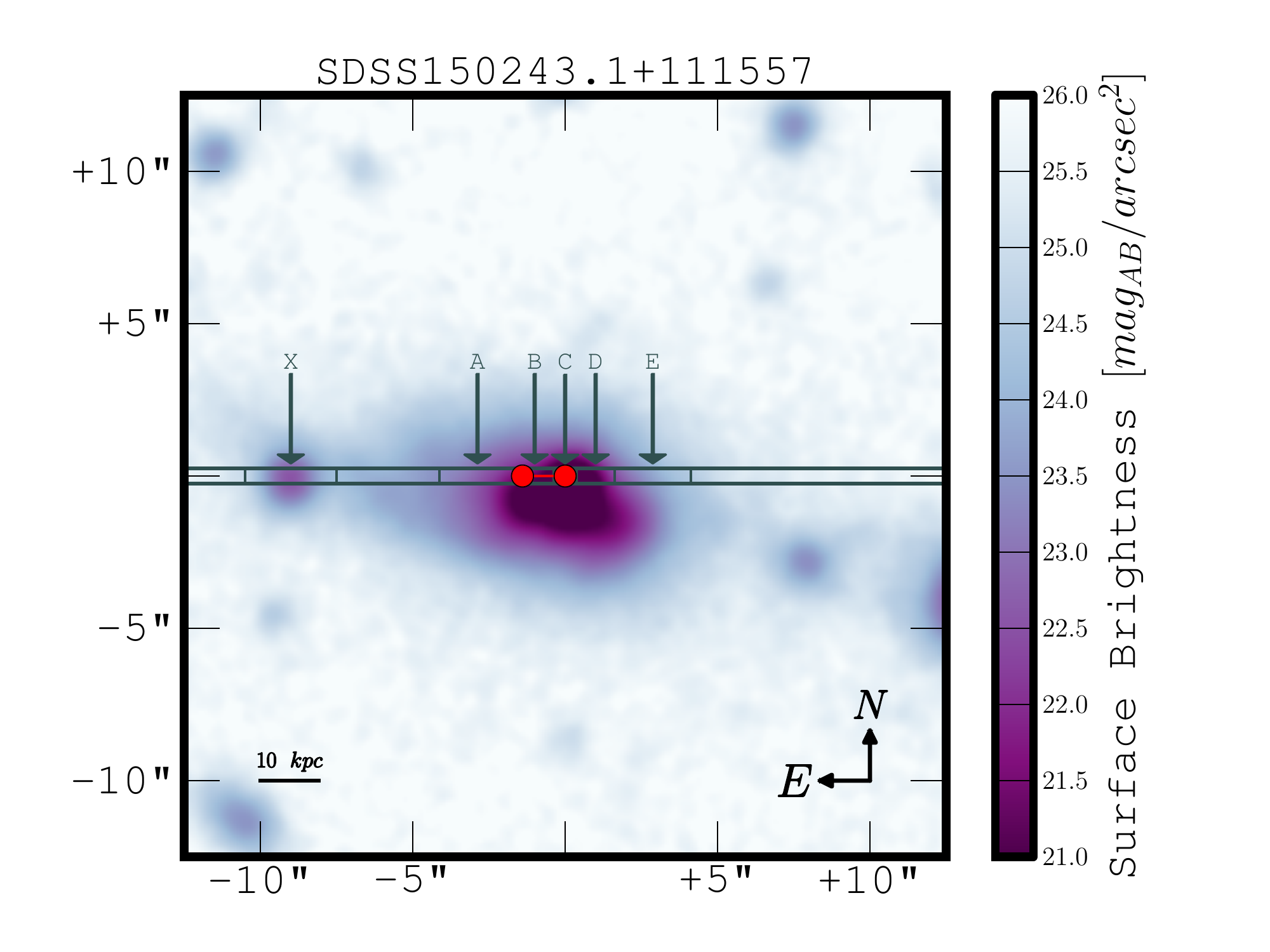}
\includegraphics[width=8.5cm]{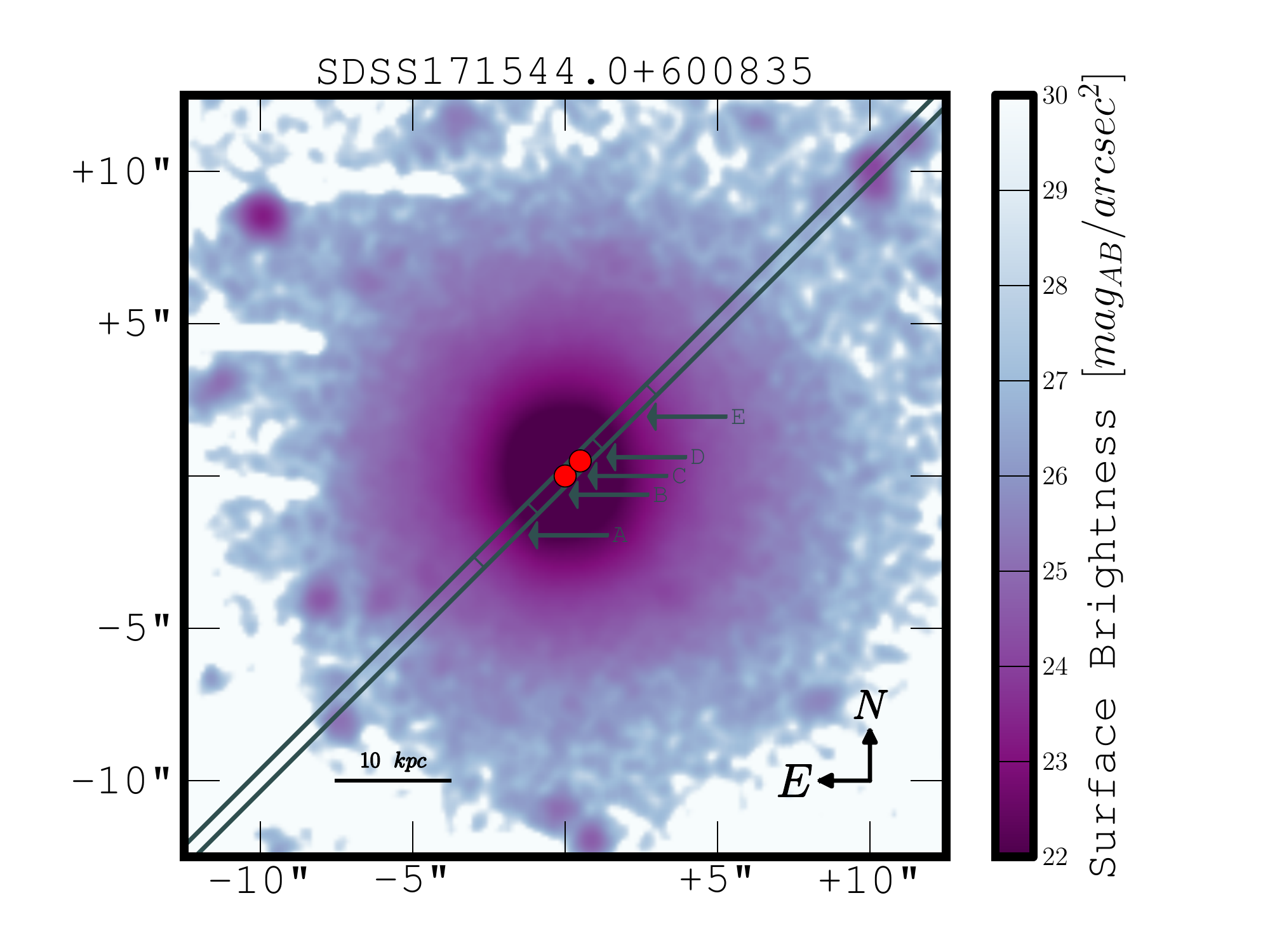}
\caption{$r$ band surface brightness maps before PSF subtraction for SDSS0952+2552 (upper left), SDSS1151+4711 (upper right), SDSS1502+1115 (lower left) and SDSS1715+6008 (lower right). Figures show 25\arcsec$\times$25\arcsec cut-outs centred on the dominant point source, the slit position as well as its approximate width are over-plotted.  North is up and east is left. Note that each plot has a slightly different scaling for the surface brightness to best show the range in each object. The separation and orientation of the two distinct cores is shown as red filled circles. This information is taken from \citet[][; SDSS0952+2552,SDSS1502+1115]{fu_mergers_2011}, \citet[][; SDSS1151+4711]{fu_nature_2012} and \citet[][; SDSS1715+6008]{comerford_chandra_2011}. In SDSS0952+2552, SDSS1151+4711 and SDSS1502+1115, the Type 1 source is located in slice C. The secondary in SDSS0952+2552 is in slices A/B, in SDSS1151+4711, it is located in Slice D and for SDSS1502+1115, it is in Slices A/B. Since no Type 1 is present in SDSS1715+6008, the locations of the two X-ray point sources are shown.
\label{F:slitmaps}}
\end{figure*}

\section{Sample}
\label{S:sample}

For simplicity, the following shortened names for objects in the sample will be used: SDSS0952+2552 (SDSS J095207.6 +255257), SDSS1151+4711 (SDSS J115106.7 +471157), SDSS1502+1115 (SDSS J150243.1 +111557) and SDSS1715+6008 (SDSS J171544.0 +600835).

The sample is selected from AGN with double-peaked $\left[ \textrm{OIII} \right]$ emission lines \citep{smith_search_2010,shen_binary_2010} at $0.1 \leqslant$ z $\leqslant 0.4$. All have double-peaked $\left[ \textrm{OIII} \right]$ lines with velocity offsets of 300 km s$^{-1}$ $\leqslant$ $\Delta$ v $\leqslant$ 750 km s$^{-1}$.

Of the four sources studied here, three (SDSS0952+2552, SDSS1151+4711, SDSS1502+1115) show two cores with similar magnitudes, separated by 4.7 - 7.4 kpc in high resolution NIR imaging, identifying them as major mergers \citep{shen_binary_2010, fu_mergers_2011}. In all three cases, Integral Field Unit (IFU) data show that each of the two NIR cores is associated with $\left[ \textrm{OIII} \right]$ emission at a different velocity and that both NIR cores have AGN spectra, verified using emission line diagnostics \citep{fu_nature_2012}.  Additionally, SDSS1502+1115 has been followed up in the radio. The data show two compact radio cores inconsistent with star formation, confirming the source as a binary AGN \citep{fu_kiloparsec-scale_2011}. We well refer to this subsample of objects as double-nucleated NIR AGN.

The fourth AGN (SDSS1715+6008) is a double-peaked $\left[ \textrm{OIII} \right]$ emitter followed up by \citet{comerford_chandra_2011} using Chandra. The data show two distinct point sources in the X-ray, identifying the object as a binary AGN. Both AGN in this source are obscured, one is possibly Compton thick. SDSS1715+6008 shows no broad line emission. A summary of sample properties can be found in Table \ref{T:objects}. We will refer to this source as an X-ray binary AGN throughout the paper.

\begin{sidewaystable}
\caption{Binary AGN sample properties and observation information. ID: SDSS name, z: redshift, $\textrm{log}(L_{[OIII] 5007} \textrm{[erg/s]})$: $\left[ \textrm{OIII} \right]$ from \citet{smith_search_2010}, $\theta [\arcsec]$: separation between AGN in arcseconds, d [kpc]: distance between AGN in kpc, Reference: reference for identification of binary AGN and measurement of AGN separation (C11: \citet{comerford_chandra_2011}; F11a: \citet{fu_mergers_2011}; F11c: \citet{fu_kiloparsec-scale_2011}; F12: \citet{fu_nature_2012}); log(R): log of radio-loudness, calculated from FIRST fluxes and SDSS $g$ band magnitudes; Observation Date (day/month/year);  seeing during observations\label{T:objects}}
\begin{tabular}{cccccccccc}
\hline
ID & Type & z  & $\textrm{log}(L_{[OIII] 5007} \textrm{[erg/s]})$ & $\theta$ [``] & d [kpc] & Reference & log(R) & Obs. Date & Seeing ($u/r/z$) [\arcsec]\\
\hline
SDSS0952+2552 & Type1/2 & 0.339 & 42.34 & 0.99 & 4.9 & F11a, F12 & $<$0.27 & 24/03/2012 &0.70/0.75/0.56\\
SDSS1151+4711 & Type1/2 & 0.318 & 43.20 & 1 & 4.7  & F12 &  0.75 & 24/03/2012 & 0.75/0.80/0.75\\
SDSS1502+1115	& Type1/2 & 0.390 & 43.35 & 1.39 & 7.4  & F11a,b, F12 & 1.8 & 19/04/2012 & 0.8/0.88/0.68 \\
SDSS1715+6008 & Type 2/2 (X-ray) & 0.157 & 42.23  & 0.68  &1.85 & C11 &  1.95 & 24/03/2012 & 0.69/0.69/0.56\\
\hline
\end{tabular}
\end{sidewaystable}

\section{Observations \& Data Reduction}
\label{S:obs}

Data were taken on the nights of March 24--25th 2012 and April 19--20th 2012 at the 10.4m Gran Telescopio Canarias (GTC) on La Palma (Canary Islands, Spain) in service mode during dark time. Observing conditions were excellent during both nights with clear sky conditions and excellent seeing (0.56-0.75\arcsec in $z$).  Observation conditions and dates are given in Table \ref{T:objects}. All observations were obtained with OSIRIS, an imager and spectrograph for the optical wavelength range, located in the Nasmyth-B focus of GTC. For all objects, both spectroscopy and imaging data were taken consecutively. All data reduction is performed using \textsc{iraf} \footnote{\textsc{iraf} is distributed by the National Optical Astronomy Observatories, which are operated by the Association of Universities for Research in Astronomy, Inc., under cooperative agreement with the National Science Foundation.} and \textsc{Pyraf}  \footnote{\textsc{Pyraf} is a product of the Space Telescope Science Institute, which is operated by AURA for NASA.}. Cosmic ray removal is performed using L.A. Cosmics \citep{van_dokkum_cosmic-ray_2001}.

Imaging data for all four objects were obtained in the Sloan $u$, $r$ and $z$ bands with exposure times of 100 s ($u$), 600 s ($r$) and 100 s ($z$), respectively. All exposures were adjusted to avoid saturation in the object and a small step dither was performed between exposures. The pixel scale is 0.125\arcsec/pix. Photometric calibrations are performed using field stars.

Spectroscopic data were obtained for all four objects with a resolution of R=2500, a slit width of 0.4\arcsec and a spectral range of 5630-7540 \AA. The chip was binned in the spatial direction, resulting in a spatial resolution of 0.25\arcsec. The data are flat-fielded and wavelength calibrated, distortions in the dispersion direction are corrected and cosmic rays removed. A 2D spectrum is then extracted, as well as 1D spectra in spatial bins. The spatial bins used were matched to the resolution of the data and are aimed at separating interesting features in the NLRs. Line fits are performed using \textsc{PySpecKit} \footnote{http://pyspeckit.bitbucket.org/html/sphinx/index.html}. Details on the line fits can be found in Appendix \ref{S:appendix}. The fit results will be discussed in Section \ref{S:results}, tables with detailed fit results as well as plots for all sources are presented in the Appendix.

2D host galaxy fits are performed using \textsc{galfit} \citep{peng_detailed_2002}. PSFs are created from field stars. The PSFs are tested by fitting them to field stars and show minimal residuals. Fits are performed in all three bands. $u$ and $z$ are primarily used to obtain galaxy magnitudes, while $r$ images are used for a detailed morphological analysis. The sky is fit locally in \textsc{galfit}. In a first step, a pure point source is performed, host galaxy components are added when needed. The position of the point source and host galaxy were not fixed with respect to each other. All fit results are shown in Table \ref{T:hosts}.

\begin{sidewaystable*}
\caption{Host galaxy fitting results. ID: object name, filter: filter in which fit was performed; $m_{AGN}$: point sources magnitude from host galaxy fit; $m_{Gal}$: fitted host galaxy magnitude, if the object is best fit by a combination of several components, the component magnitudes are added; $r_{gal}$ [\arcsec/kpc]: host galaxy effective radius in pixels/kpc respectively; sersic: best fit sersic index, if fits with free sersic indexes diverge, D is given if a disk (sersic index=1) provides the better fit and B if a deVacoleurs profile (sersic index=4) fits the data better, B/D is given if the best fit is a disk/bulge combination; $M_{AGN}$: absolute AGN magnitude, $M_{GAL}$: absolute galaxy magnitude; comments: comments on morphology and fit.}
\begin{tabular}{cccccccccc}
\hline
ID & Filter & $m_{AGN}$ & $m_{Gal}$ & $r_{gal}$ [\arcsec] & $r_{gal}$ [kpc] & Sersic & $M_{AGN}$ & $M_{Gal}$ & Comments \\
\hline
SDSS0952+2552 & $u$ & 18.82 $\pm$ 0.07 & 23.29 $\pm$ 1.70 & 0.3 & 1.5 & D & -22.44  & -17.97 & --\\
SDSS0952+2552 & $r$ & 18.35 $\pm$ 0.11 & 19.33 $\pm$ 0.11 & 0.9 & 4.4 & D (1.2) & -22.91 & -21.93 &  double core, weak tidal tails\\
SDSS0952+2552 & $z$ & 17.66 $\pm$ 0.13 & 18.54 $\pm$ 0.13 & 0.9 & 4.4 & D (1.4) & -23.6 & -22.72 & -- \\
 & & & & & & & & & \\
SDSS1151+4711 & $u$ & 17.74 $\pm$ 0.03 & -- & -- & -- &  -- & -23.36 &  -- & host galaxy unresolved \\
SDSS1151+4711 & $r$ & 15.91 $\pm$ 0.07 & 17.66 $\pm$ 0.07 & 1.4 & 6.52 & D & -25.19 & -23.44 & strong tidal tails\\
SDSS1151+4711 & $z$ & 15.38 $\pm$ 0.02 & 16.64 $\pm$ 0.03 & 0.4 & 1.9 & D & -25.72 & -24.46 & --\\
 & & & & & & & & & \\
SDSS1502+1115 & $u$ & 19.57 $\pm$ 0.12 & 21.03 $\pm$ 0.43 & 0.2 & 1.1 & D & -22.04 & -20.59 & --\\
SDSS1502+1115 & $r$ & 18.14 $\pm$ 0.10 & 17.42 $\pm$ 0.10 & ? & ? & B/D & -23.44 & -24.20 & disturbed morphology \\
SDSS1502+1115 & $z$ & 17.23 $\pm$ 0.10 & 16.65 $\pm$ 0.11 & ? & ? & B/D & -24.39 & -24.97 & -- \\
 & & & & & & & \\
SDSS1715+6008 & $u$ & 21.62 $\pm$ 0.19 & 20.70 $\pm$ 0.21 & 0.6 & 1.6 & ? & -18.11 & -18.67 & -- \\
SDSS1715+6008 & $r$ & 18.81 $\pm$ 0.04 & 16.86 $\pm$ 0.04 & 0.7/1.1 & 1.9/3 & B/D & -21.27 & -22.61 & -- \\
SDSS1715+6008 & $z$ & 18.99 $\pm$ 0.05 & 16.54 $\pm$ 0.51 & 1.2/1.8& 3.3/4.9 & B/D & -20.38 & -22.83 & -- \\
\hline
\end{tabular}
\label{T:hosts}
\end{sidewaystable*}

\section{Results}
\label{S:results}

In this section we present the results for the sample of four AGN. Detailed properties of individual sources are discussed in Appendix \ref{S:appendix}.

\subsection{Imaging: Merger Stages}
\label{S:imaging}

The $r$-band imaging data are shown in Fig. \ref{F:slitmaps} with the approximate slit positions overlaid.  The three objects with double-nucleated IR structure (SDSS0952+2552, SDSS1151+4711, SDSS1502+1115) show clear merger features. In all cases, we resolve the double-nuclei detected in the NIR. SDSS0952+2552 and SDSS1151+4711 both show extended low surface brightness emission on scales $\gtrsim 10$ kpc. SDSS1502+1115 on the other hand, while clearly disturbed and showing several distinct knots of emission in $r$ is compact and shows only weak tidal features. Note that this source shows the largest separation between the nuclei, indicating an early merger stage. However, it should be noted that these three sources were chosen to have double-nuclei, clear merger features are therefore expected. The X-ray binary AGN SDSS1715+6008 stands out since it shows no signs of merging or disturbance down to very low surface brightnesses ($\sim$ 27 mag/arcsec$^{2}$ in $r$). The structures seen in the outskirts of the galaxy are consistent with background sources.

Both SDSS0952+2552 and SDSS1151+4711 are best fit by a disk component plus point source, though with considerable residuals due to the disturbed nature. A multi-component fit is required for SDSS1502+1115 as well as SDSS1715+6008, both are best fit by a point source as well as both a disk and bulge component. Details of the host galaxy fits are provided in Table \ref{T:hosts}.

An interesting finding of our deep imaging study is that out of the four objects studied, all but SDSS1715+6008 have close faint companions. In the case of SDSS0952+2552 and SDSS1151+4711 the companions are faint disk galaxies with separations of $\sim$ 50 kpc, connected to the main galaxy through extended emission. In the case of SDSS1502+1115, the separation is somewhat larger ($\sim$100 kpc) and the companion is unresolved. The slit covers the companions for all three objects, and for two objects either absorption or emission at the redshift of the main object is detected. This will be discussed in more detail in Section \ref{S:spec}.

In summary, all but one of the objects studied show clear signs of interaction with strong tidal tails and disturbed morphologies, consistent with ongoing mergers. The remaining object, which is an X-ray binary AGN with considerably smaller separation than the double-nuclei AGN ($\sim$2 kpc, compared to $\sim$5-8 kpc for the double-nuclei sources) shows an undisturbed morphology even at the low surface brightnesses reached. The smaller separation in this source suggests a later merger stage, broadly consistent with the less disturbed morphology.

The 2D fits in $u/r/z$ provide integrated and spatially resolved information on the host galaxy colors. The $u-r$ colors place the galaxies in the red sequence \citep[e.g.][]{strateva_color_2001,balogh_bimodal_2004,schawinski_galaxy_2010}, consistent with quenched star formation. However, dust reddening can cause similarly red $u-r$ colors \citep[e.g.][]{cardamone_dust-corrected_2010}.

\begin{figure*}
\includegraphics[width=8cm]{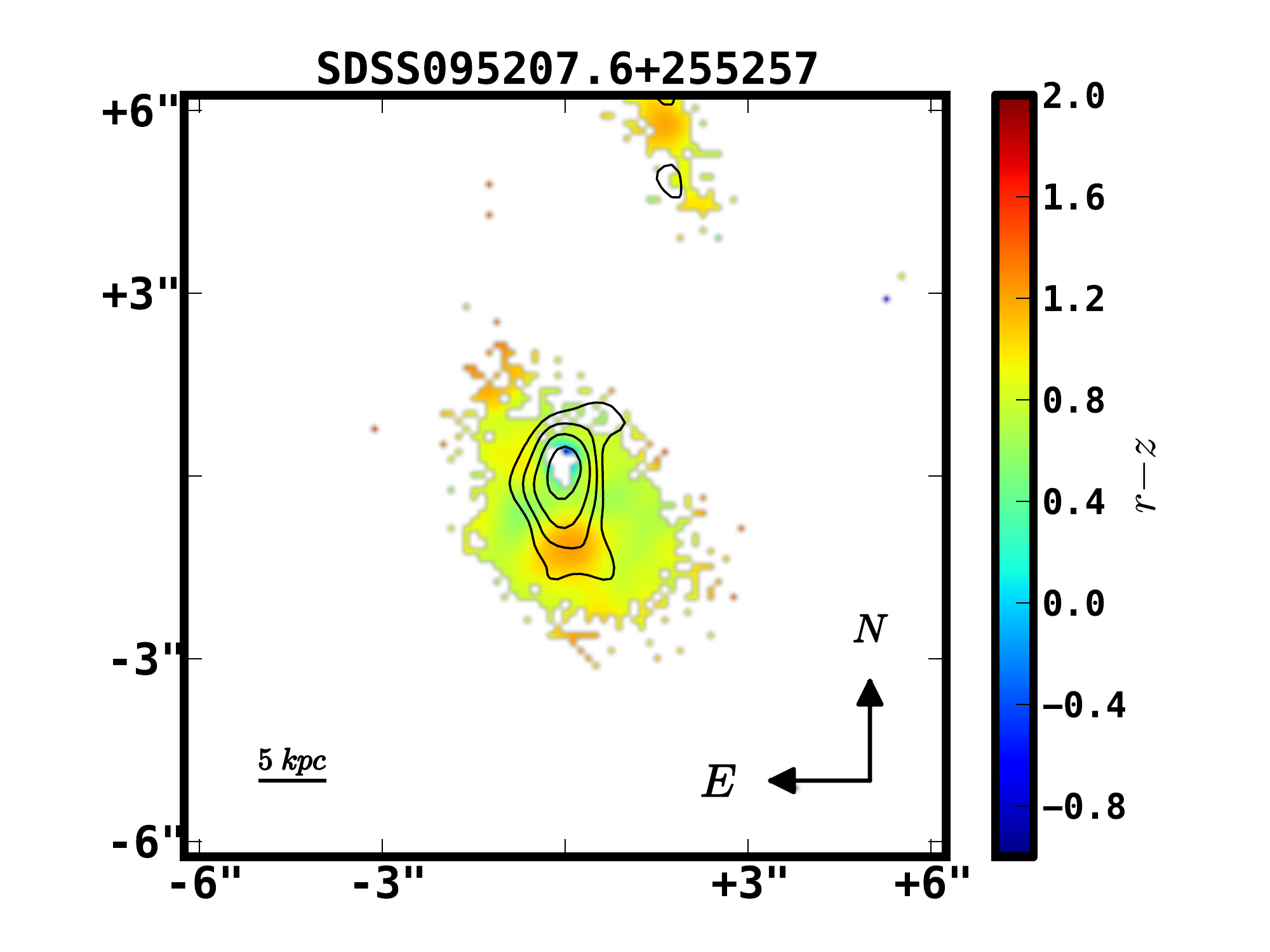}
\includegraphics[width=8cm]{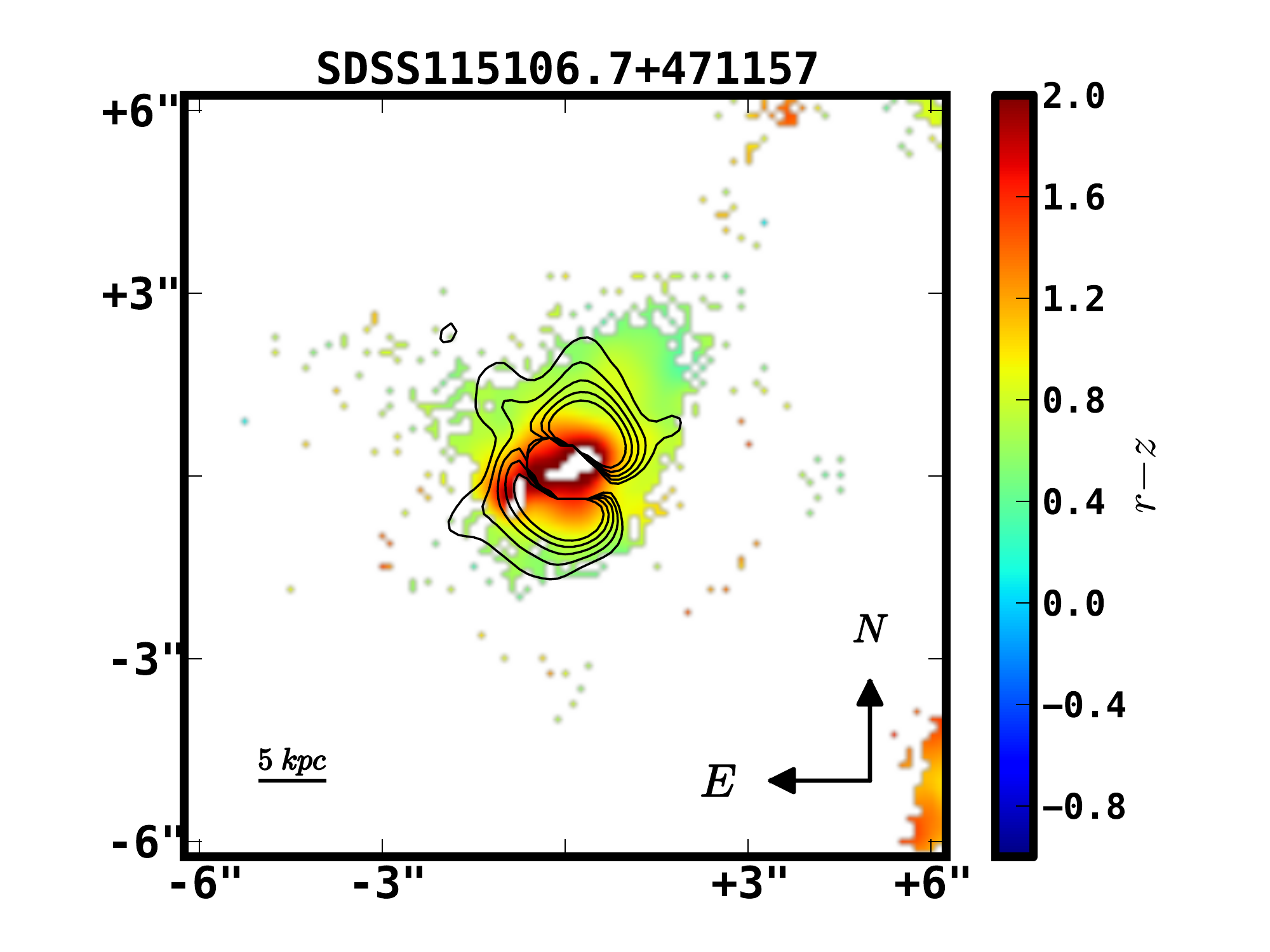}
\includegraphics[width=8cm]{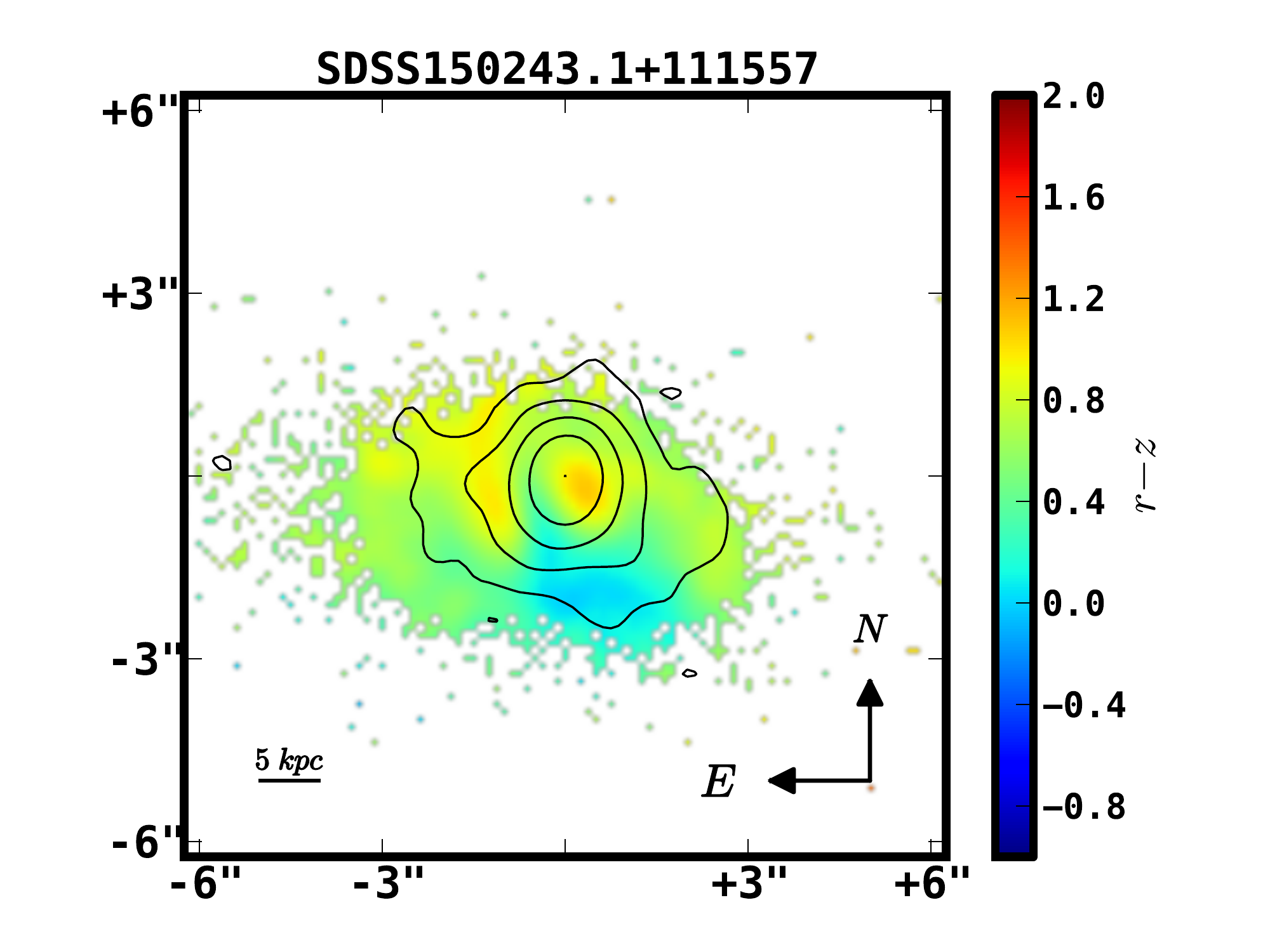}
\includegraphics[width=8cm]{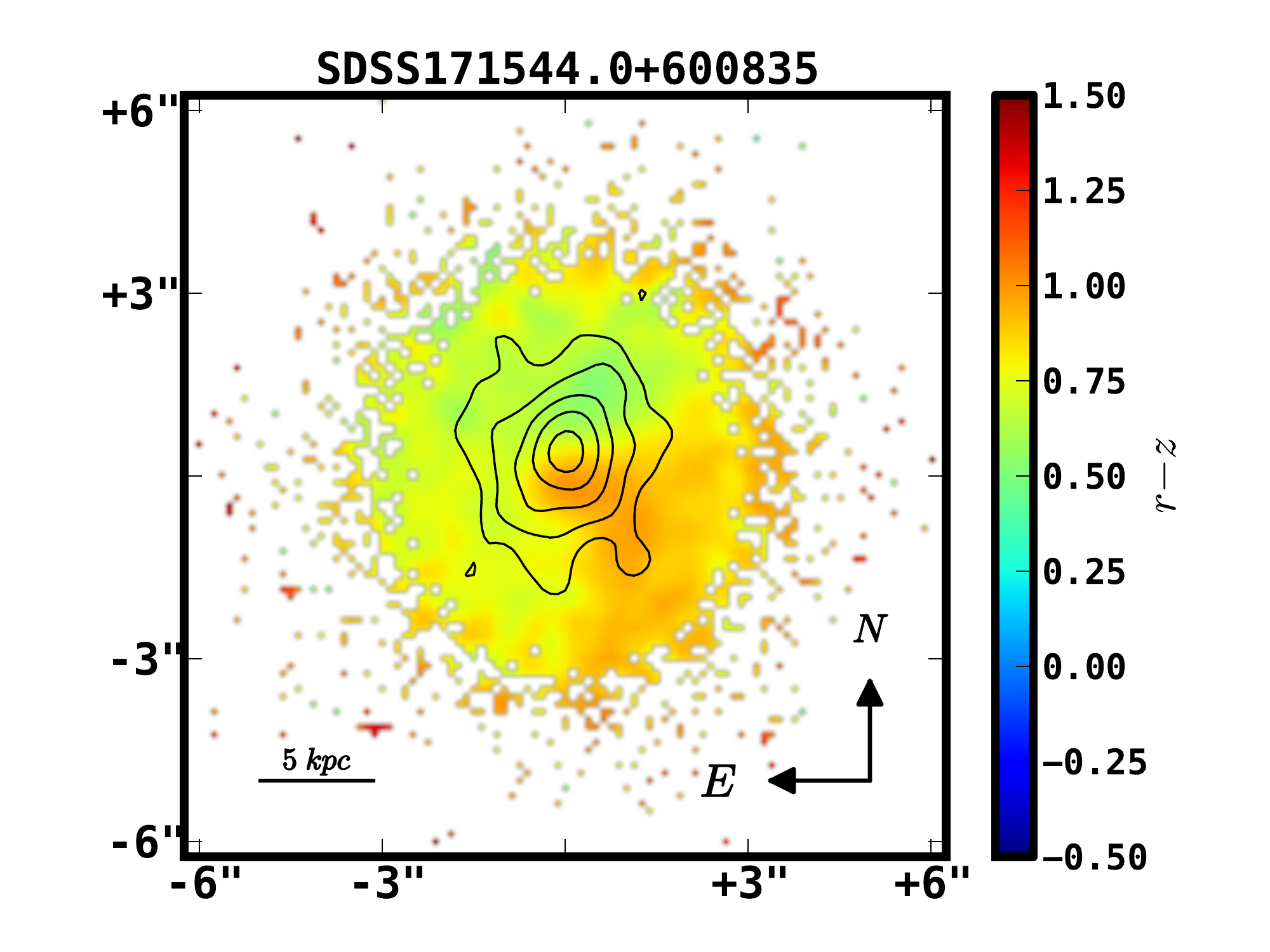}
\caption{Color structure of host galaxy (after subtraction of point source). The image color shows $r-z$, both $r$ and $z$ band images are smoothed using a Savitzky-Golay algorithm with a window size of $3\times3$ pixels, an order of one and three iterations. The contours show the $u$ surface brightness with linearly spaced levels from 22 -- 23.5 mag/arcsec$^{2}$. North is up and east is left.
\label{F:colourmaps}}
\end{figure*}

Spatially resolved colors are shown in Figure \ref{F:colourmaps}. SDSS0952+2552, SDSS1151+4711 and SDSS1502+1115 all show redder $r-z$ colors towards the central peaks, the $u$ band emission appears to be centred on the areas with redder $r-z$ colors. The redder colors can be interpreted both as stronger reddening towards the central region as well as strong H$\alpha$ emission since H$\alpha$ falls in the $z$ band  at the redshifts of these three objects. Combined with the co-spatial $u$ band emission tracing younger stellar populations, the color maps are consistent with central compact ($<$5 kpc) starbursts. Note that in the case of SDSS1151+4711 and SDSS1502+1115, due to the disturbed morphology, artefacts from the PSF fit were visible in the most central pixels. Additionally, the extreme reddening seen in the very center of SDSS1151+4711 might be an indication that the point source has been slightly over-subtracted due to the complex nature of the profile fit of the merging galaxies.

SDSS1715+6008 shows a color structure quite different from the other sources. The $u$ band emission is extended and without clear structure while the $r-z$ color map shows a single funnel-shaped area of almost one magnitude redder emission in the south-west corner of the galaxy. Due to the lower redshift $\left[ \textrm{OIII} \right]$ falls just in the blue edge of the $r$-band for SDSS1715+6008. The reddened wedge could therefore either be interpreted as an area with higher extinction or a region where $\left[ \textrm{OIII} \right]$ emission is suppressed. Unfortunately, the slit placement was almost exactly orthogonal to the wedge feature and therefore the strength of $\left[ \textrm{OIII} \right]$ in the reddened area cannot be assessed.

\subsection{Spectroscopy: Kinematics}
\label{S:spec}

\begin{figure*}
\begin{center}
\includegraphics[width=18cm]{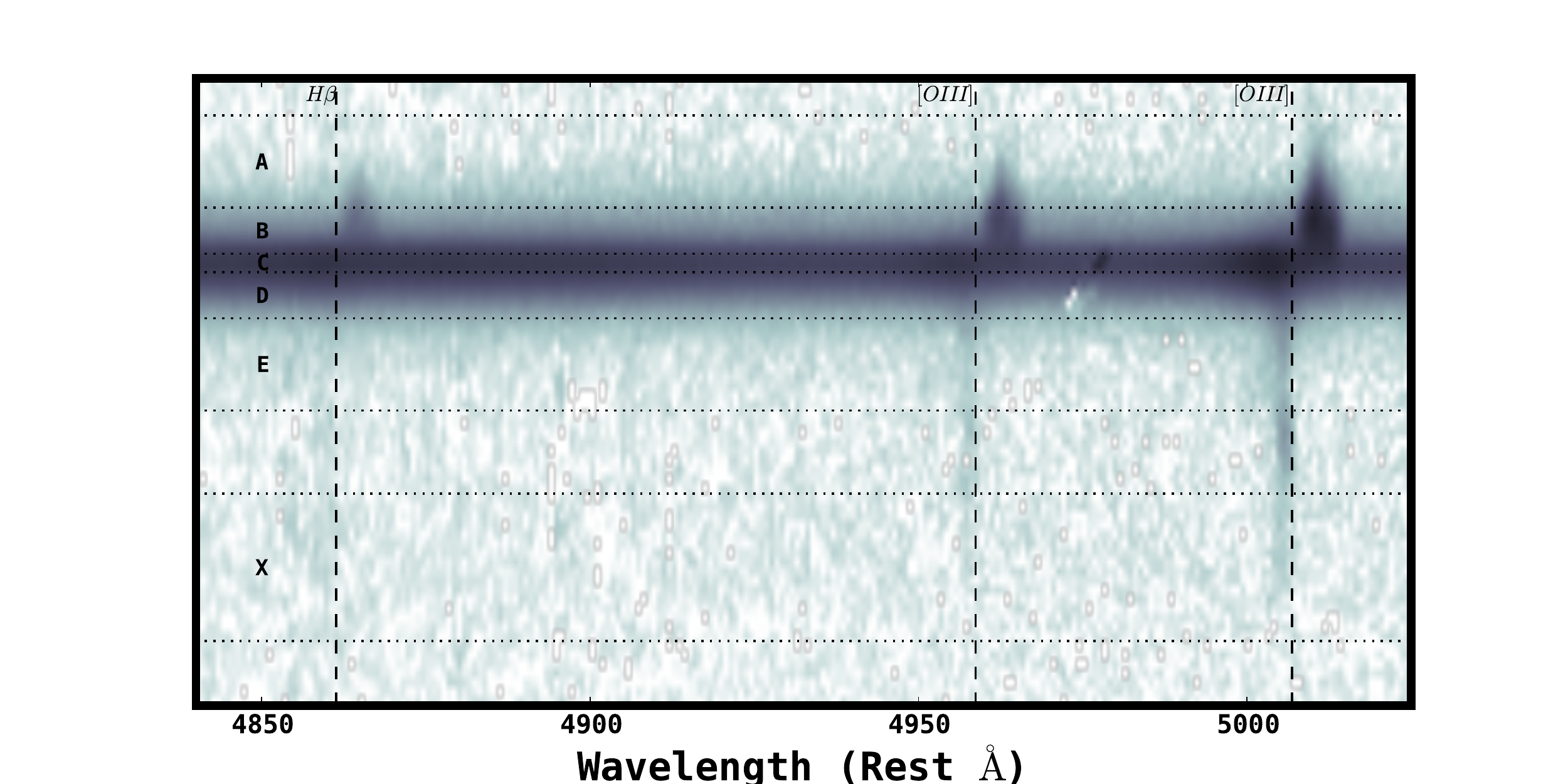}
\caption{Two dimensional spectrum of SDSS0952+2552 showing H$\beta$ and the $\left[ \textrm{OIII} \right]$ doublet. The image is shown in log scale, slices are as shown in Fig. \ref{F:slitmaps}. Vertical dashed lines show the wavelengths of H$\beta$ and the $\left[ \textrm{OIII} \right]$ doublet. A narrow line width with a velocity close to systemic is visible through slices D-X. A separate broad line width component red-shifted from systemic can be seen in slices A and B, where the suspected secondary AGN is located.
\label{F:twod_0952}}
\end{center}
\end{figure*}

\begin{figure*}
\begin{center}
\includegraphics[width=18cm]{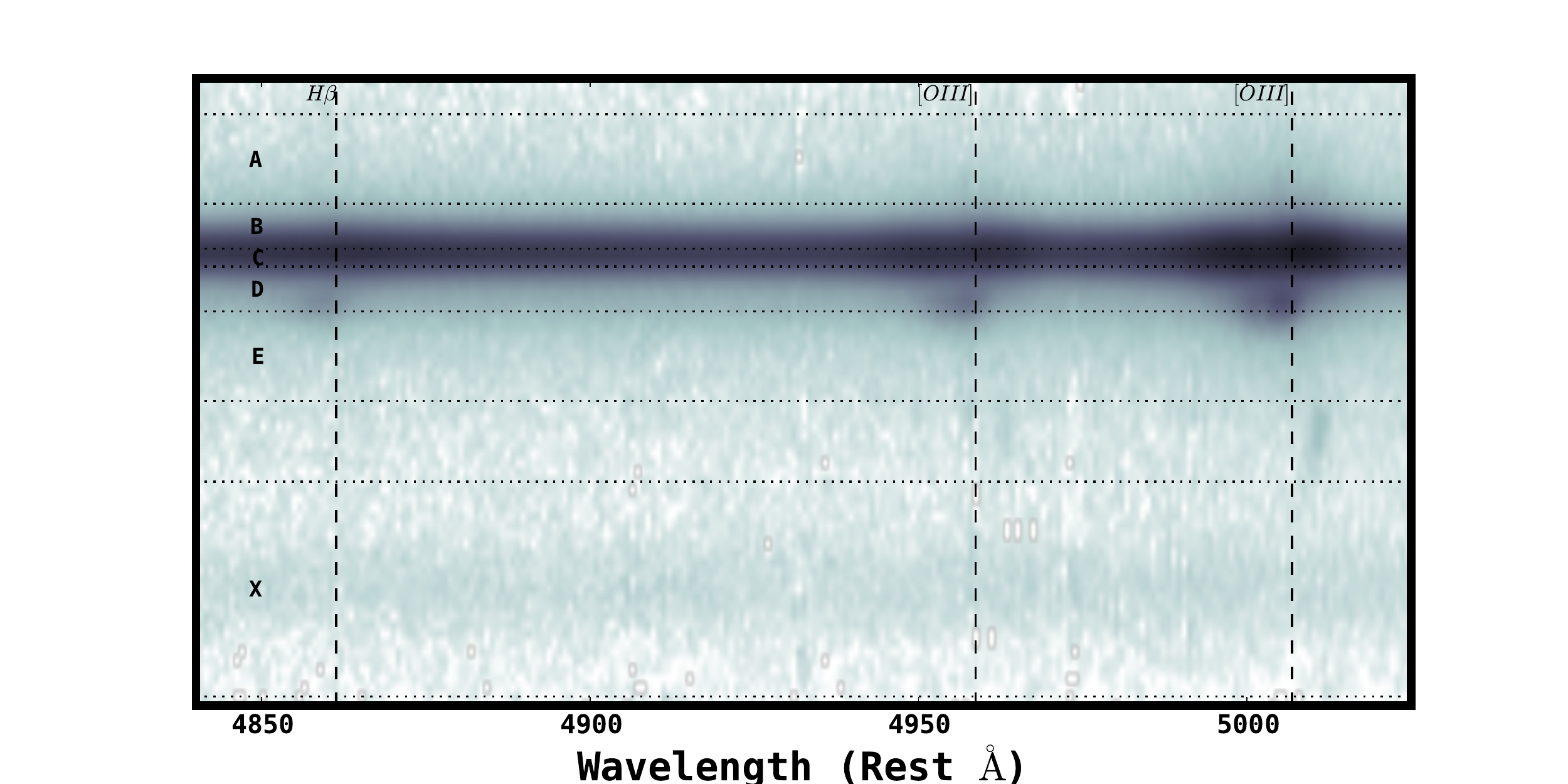}
\caption{Two dimensional spectrum of SDSS1151+4711 showing H$\beta$ and the $\left[ \textrm{OIII} \right]$ doublet. The image is shown in log scale, slices are as shown in Fig. \ref{F:slitmaps}. Vertical dashed lines show the wavelengths of H$\beta$ and the $\left[ \textrm{OIII} \right]$ doublet. The 2D spectrum shows two spatially distinct $\left[ \textrm{OIII} \right]$ regions only marginally separated in velocity space. The suspected secondary AGN is located in Slice D.
\label{F:twod_1151}}
\end{center}
\end{figure*}

\begin{figure*}
\begin{center}
\includegraphics[width=18cm]{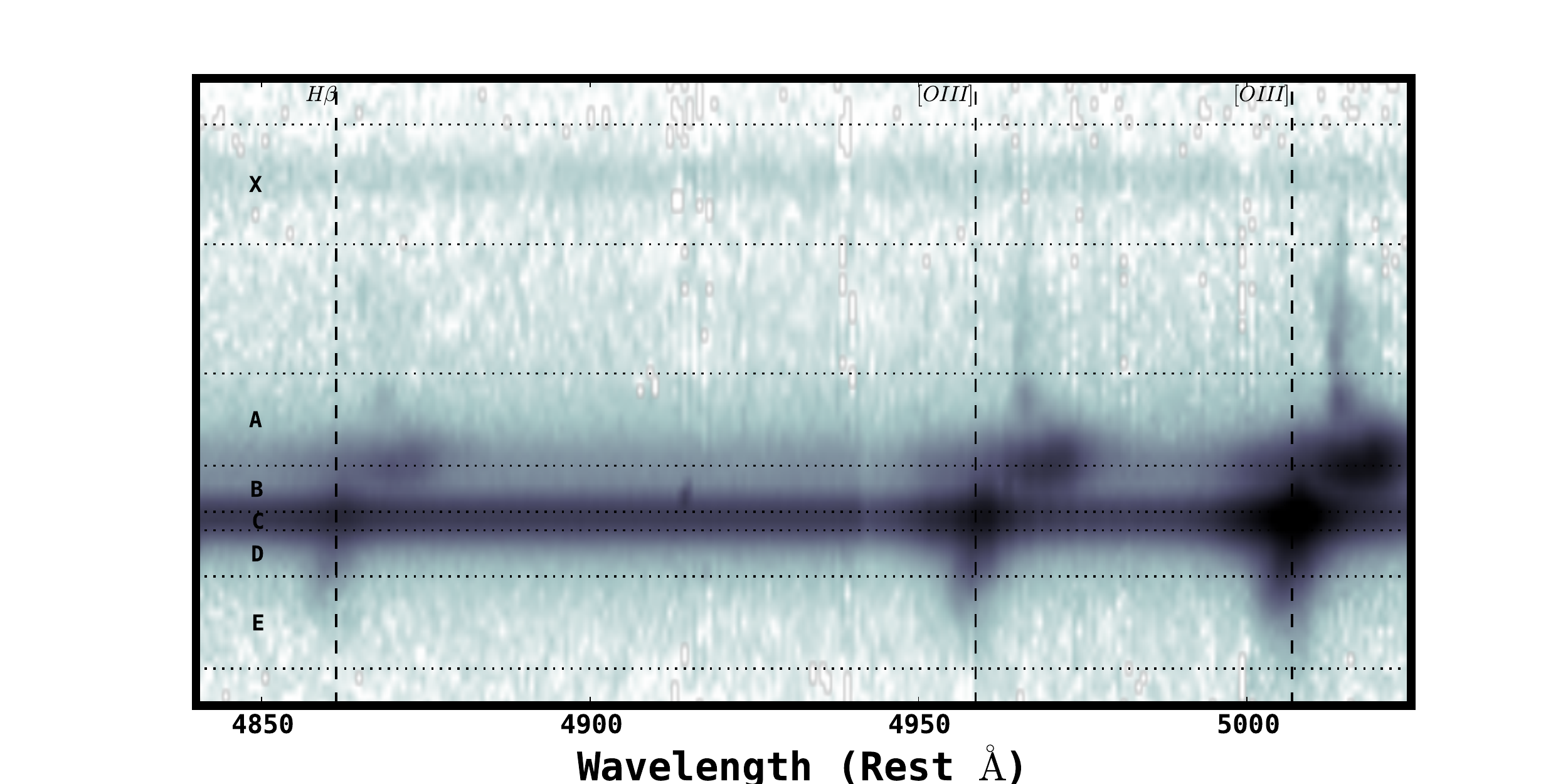}
\caption{Two dimensional spectrum of SDSS1502+1115 showing H$\beta$ and the $\left[ \textrm{OIII} \right]$ doublet. The image is shown in log scale, slices are as shown in Fig. \ref{F:slitmaps}. Vertical dashed lines show the wavelengths of H$\beta$ and the $\left[ \textrm{OIII} \right]$ doublet. The 2D spectrum is very complex. Two distinct strong $\left[ \textrm{OIII} \right]$  regions associated with the two AGN are visible, one in Slice C, the other in Slices A/B. The secondary AGN in slice A/B is a Type 2 AGN, red-shifted with respect to the primary. A narrow line width $\left[ \textrm{OIII} \right]$ region is seen emerging from the secondary AGN, blue-shifted from the secondary, but red-shifted from systemic. A broad line width $\left[ \textrm{OIII} \right]$ region blue-shifted from the primary is also seen in slices D/E.
\label{F:twod_1502}}
\end{center}
\end{figure*}

\begin{figure*}
\begin{center}
\includegraphics[width=18cm]{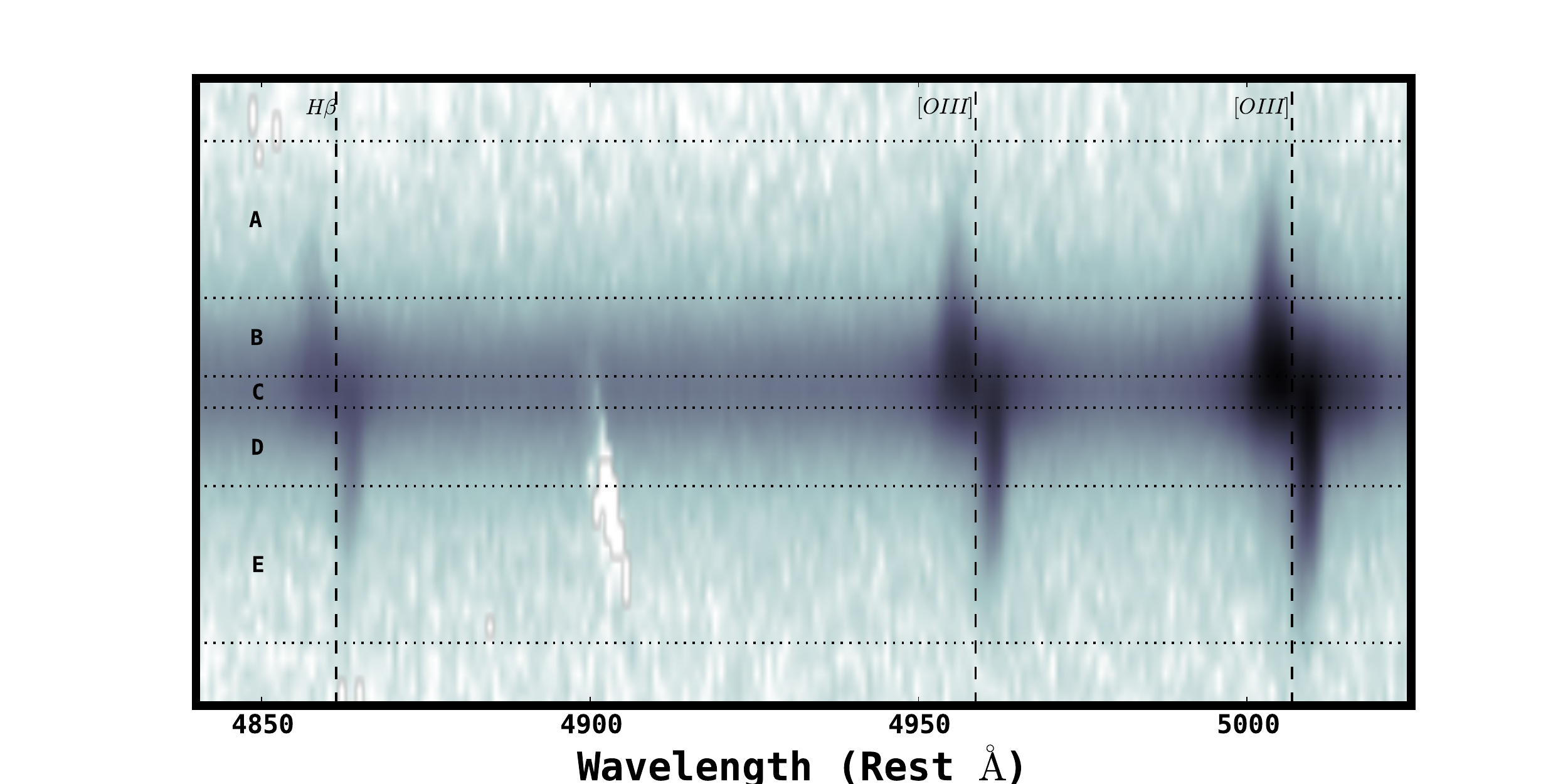}
\caption{Two dimensional spectrum of SDSS1715+6008 showing H$\beta$ and the $\left[ \textrm{OIII} \right]$ doublet. The image is shown in log scale, slices are as shown in Fig. \ref{F:slitmaps}. Vertical dashed lines show the wavelengths of H$\beta$ and the $\left[ \textrm{OIII} \right]$ doublet. The 2D spectrum shows blue-shifted emission in Slices A/B and red-shifted emission in slices D/E.
\label{F:twod_1715}}
\end{center}
\end{figure*}

\begin{figure*}
\begin{center}
\includegraphics[width=8cm]{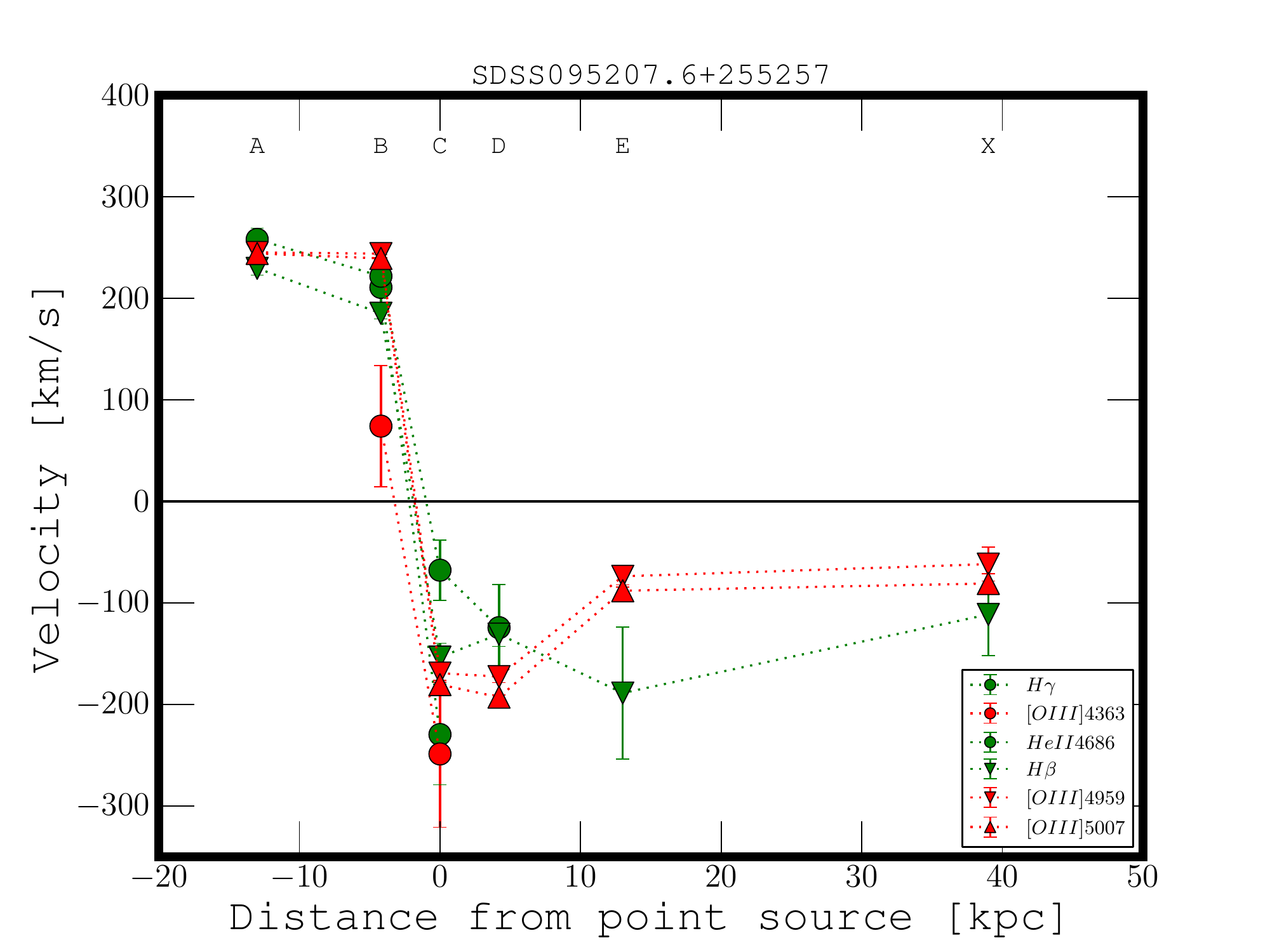}
\includegraphics[width=8cm]{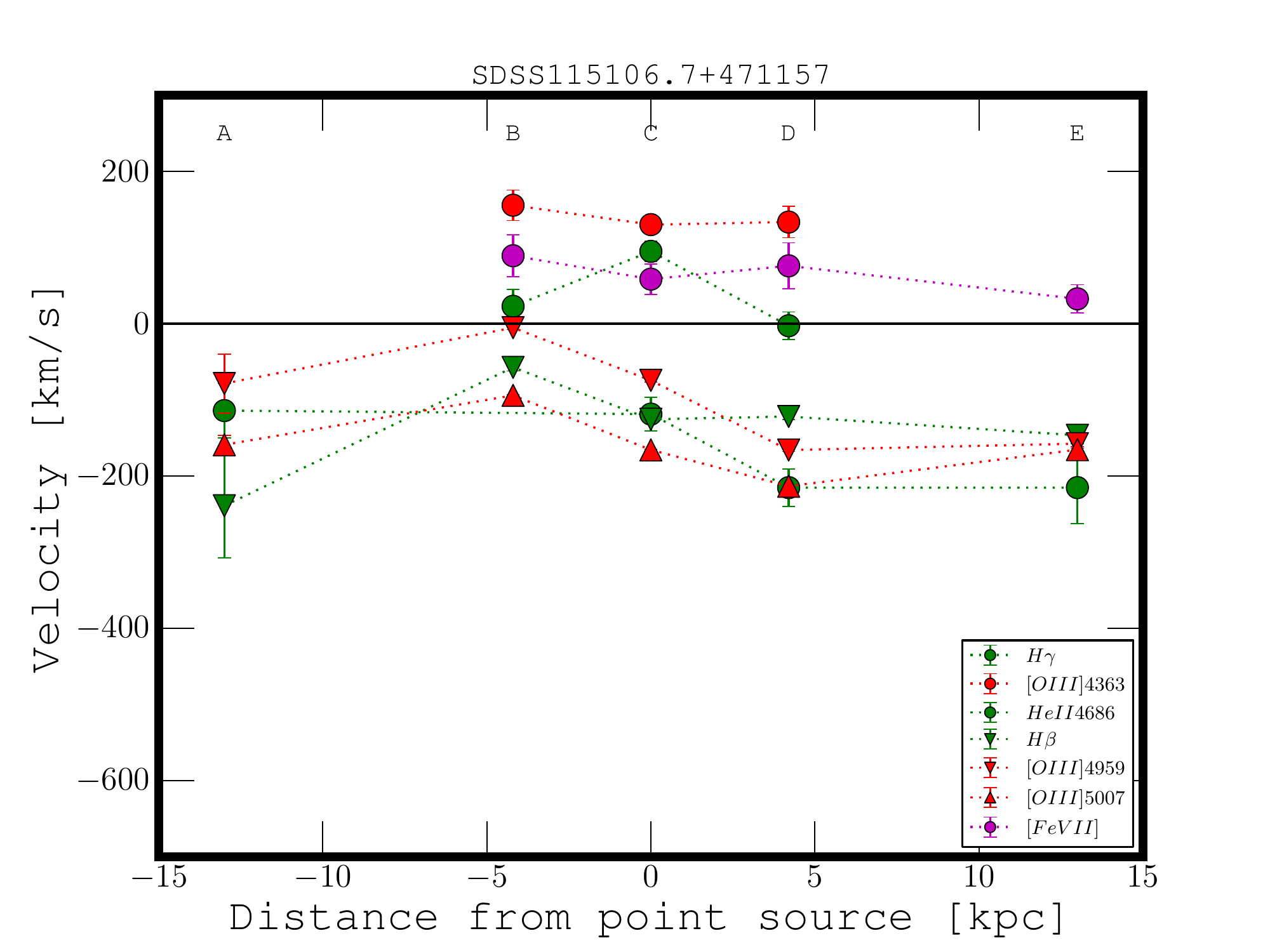}
\includegraphics[width=8cm]{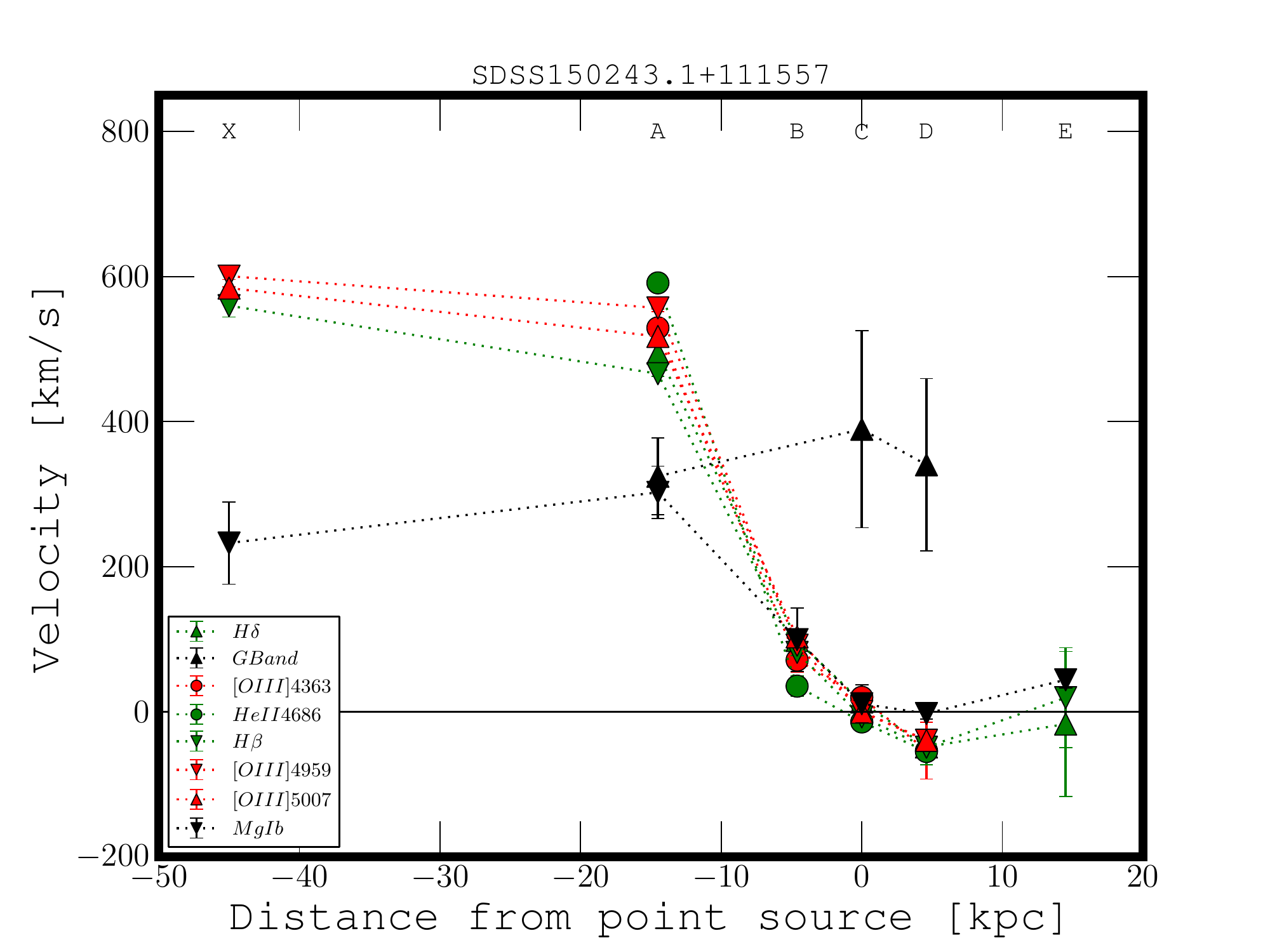}
\includegraphics[width=8cm]{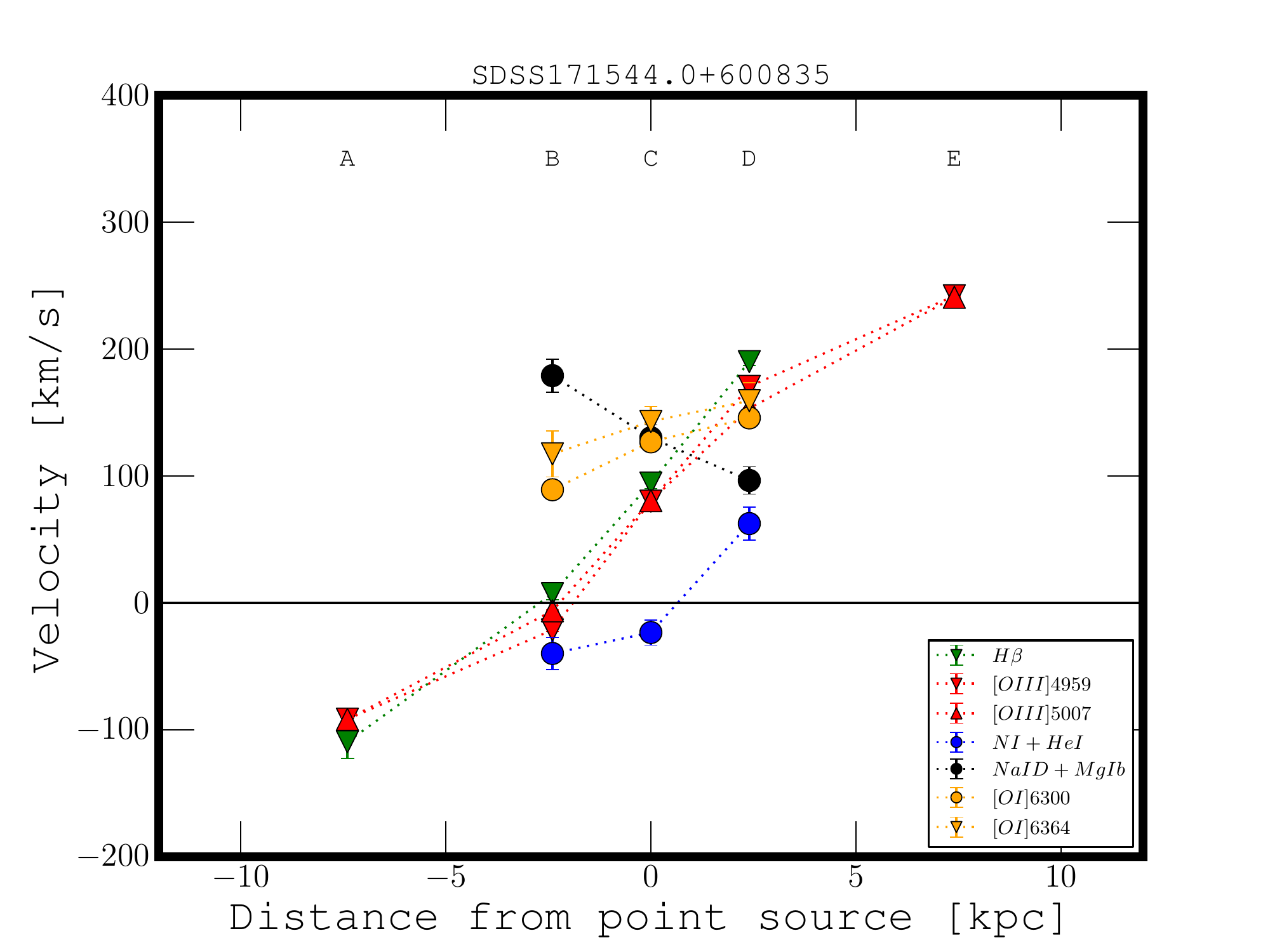}
\caption{Kinematic structure of different lines in all objects, the location of the different slices within the galaxy are labelled with letters and are shown in Figure \ref{F:slitmaps}. The velocities are calculated with respect to the catalogued host galaxy redshifts from \citet{smith_search_2010}. For details about the parameters used in the line fits, see Appendix \ref{S:appendix}.
\label{F:kinematics}}
\end{center}
\end{figure*}

Long-slit spectra provide detailed spatially resolved information about gas and stellar kinematics. In all plots and analysis we use redshifts from \citet{smith_search_2010}. When available, we use redshifts determined from stellar absorption lines, for the two cases where no host galaxy redshift could be determined (SDSS1151+4711, SDSS1502+1115), we use the redshift of the $\left[ \textrm{OIII} \right]$ emission, in particular, the higher redshift $\left[ \textrm{OIII} \right]$ line is used. 

Figures \ref{F:twod_0952} - \ref{F:twod_1715} show the 2D spectra for our AGN sample covering H$\beta$ and the $\left[ \textrm{OIII} \right]$ 5007/4969 doublet. Additionally, detailed line fits are performed, as described in Section \ref{S:obs}. All details of the fits are summarized in the Appendix and in Tables \ref{T:lines_0952}-\ref{T:lines_1715b} as well as Figures \ref{F:lines_0952},\ref{F:lines_1151},\ref{F:lines_1502},\ref{F:lines_1715} (detailed maps of individual lines in each spatial slice) and Figures \ref{F:spec_0952},\ref{F:spec_1151},\ref{F:spec_1502},\ref{F:spec_1715} (profiles of $\left[ \textrm{OIII} \right]$ doublet and H$\beta$ in different spatial slices). From these fits, we derive the kinematics of the gas and stars along the slit (see Fig. \ref{F:kinematics}). Our spatial resolution corresponds to $\sim$ 1-5 kpc, depending on object.

Clearly extended $\left[ \textrm{OIII} \right]$ emission, out to scales of $\sim$ 50-100 kpc is observed in all objects in the sample (see Fig. \ref{F:twod_0952}-\ref{F:twod_1715}). Clear clumps in the NLR are apparent in the three sources with NIR double nuclei (see Fig. \ref{F:twod_0952}-\ref{F:twod_1715}). In two cases, the clumps are unresolved at the spatial resolution of our data ($\sim$1-5 kpc, Fig. \ref{F:twod_1151}; SDSS1151+4711 and Fig. \ref{F:twod_1502}; SDSS1502+1115). These NLR sizes are consistent with those of AGN with similar $\left[ \textrm{OIII} \right]$  luminosities \citep{bennert_size_2002,schmitt_hubble_2003,bennert_size_2006,fu_extended_2009,humphrey_integral-field_2010,greene_feedback_2011,liu_observations_2013}, while the overall NLR sizes are comparable to AGN two orders of magnitude brighter \citep{liu_observations_2013}.

The spatial and kinematic structure of the $\left[ \textrm{OIII} \right]$ features in our sample differ quite significantly between sources. Amongst all sources, a range of kinematic structures is observed, including very spatially extended $\left[ \textrm{OIII} \right]$ emission with narrow line width and no velocity gradients (seen in SDSS0952+2552; see Fig. \ref{F:twod_0952}, \ref{F:lines_0952}, \ref{F:spec_0952}, and SDSS1502+1115; see Fig. \ref{F:twod_0952}, \ref{F:lines_0952}, \ref{F:spec_0952}), compact $\left[ \textrm{OIII} \right]$ emission that is spatially un- or marginally resolved (seen in SDSS1151+4711; Fig. \ref{F:twod_1151}, and SDSS1502+1115; Fig. \ref{F:twod_1502}) as well as extended $\left[ \textrm{OIII} \right]$ emission with substantial line widths and gradients (seen in SDSS0952+2552; Fig. \ref{F:twod_0952}, SDSS1502+1115; Fig. \ref{F:twod_1502}, and SDSS1715+6008; Fig. \ref{F:twod_1715}). Many of the objects show several distinct features, indicating a NLR consisting of spatially separated components. In SDSS0952+2552, SDSS1151+4711 and SDSS1502+1115 broad wings in $\left[ \textrm{OIII} \right]$ with velocities up to $\sim$ 1500 km s$^{-1}$ are detected (Figures \ref{F:spec_0952}, \ref{F:spec_1151}, \ref{F:spec_1502}).

In SDSS0952+2552 (Fig. \ref{F:twod_0952}), two $\left[ \textrm{OIII} \right]$ clouds extend from the center, the southern (Slice A) part shows red-shifted $\left[ \textrm{OIII} \right]$ emission, while the northern (Slice E) part shows narrow slightly blue-shifted emission tracing along the bridge that connects the source with its nearby companion. Both features have line widths $\sim$200 km s$^{-1}$, consistent with significant internal broadening. The extent of this blue-shifted feature is quite remarkable since it spans about 50 kpc with little or no velocity gradient. $\left[ \textrm{OIII} \right]$ is also detected in the neighbouring galaxy. The two components are kinematically separated by $\sim500$ km s$^{-1}$ (Fig. \ref{F:kinematics}). The primary Type 1 AGN is at the same velocity as the blue-shifted feature, while the suspected secondary is associated with the red-shifted component. While the two components are not spatially separated, they appear to be physically separate features given the abrupt change in velocity, there is no smooth transition in velocity between the two features, but rather an overlap, indicating that the NLR consists of spatially distinct clumps. At the location of the continuum source, $\left[ \textrm{OIII} \right]$ shows strong broad wings out to $>1000$ km s$^{-1}$ with respect to systemic, indicative of an outflow.

In SDSS1151+4711 (Fig. \ref{F:twod_1151}), two very compact $\left[ \textrm{OIII} \right]$ components are detected, one associated with the Type 1 AGN, the other at the location of the suspected secondary Type 2 AGN. The two components are not connected and appear as distinct clumps. Kinematically, the two $\left[ \textrm{OIII} \right]$ components are not separated (Fig. \ref{F:kinematics}). The $\left[ \textrm{OIII} \right]$ profiles shows a marginally separated component in the blue wing of the line. The general $\left[ \textrm{OIII} \right]$ profile is broad, with substantial red wings (out to $\sim$1000 km s$^{-1}$ in the central region), and even broader blue wings ($>$1500 km s$^{-1}$). HeII 4686 and $\left[ \textrm{FeVII} \right]5721$ are both detected in this source, but similar to the other lines, no significant change in profile or velocity along the slit is observed (see Fig. \ref{F:lines_1151}). 

SDSS1502+1115 shows the most complex kinematic structure (Fig. \ref{F:twod_1502}). This source is a confirmed radio binary AGN \citep{fu_kiloparsec-scale_2011}. There are two very distinct cores of $\left[ \textrm{OIII} \right]$ and continuum emission associated with the two AGN. Additionally, there are other extended $\left[ \textrm{OIII} \right]$ components. The $\left[ \textrm{OIII} \right]$ emission in the Type 1 AGN (in Slice C, Fig. \ref{F:twod_1502}), extends out to about 10 kpc, with large line width ($\sim$800 km s$^{-1}$) and larger blue-shifts further from the core, (100 km s$^{-1}$ in C to -40 km s$^{-1}$ in E), as can be seen in Figures \ref{F:lines_1502} and \ref{F:spec_1502}. The gas in the NLR associated with this AGN appears to trace the stellar kinematics. The compact $\left[ \textrm{OIII} \right]$ region associated with the Type 2 secondary AGN (in Slice A/B, Fig. \ref{F:twod_1502}) is red-shifted with respect to the primary AGN by about $\sim$500 km s$^{-1}$. The $\left[ \textrm{OIII} \right]$ velocity traces that of the stellar component (see Fig. \ref{F:kinematics}). $\left[ \textrm{OIII} \right]$ emission with a velocity dispersion $\leq$ 100 km s$^{-1}$ extends from the secondary AGN to the associated object in Slice X, stretching out for close to 100 kpc. It is blue-shifted from the secondary, but red-shifted from the primary AGN. The overall kinematics are too complex to isolate the feature in the collapsed spectra. The compact companion in the east shows stellar absorption at systemic velocity. The general $\left[ \textrm{OIII} \right]$ profile is extremely broad in this object, both components have strong wings, overall, the $\left[ \textrm{OIII} \right]$ emission spans from $\sim$-1000 to +1500 km s$^{-1}$ with respect to systemic. HeII is detected throughout the galaxy, following $\left[ \textrm{OIII} \right]$ in velocity (see Figures \ref{F:kinematics}, \ref{F:lines_1502}). Due to the lower signal-to-noise ratio in this line, it is unclear if it shows the same extreme velocity wings as $\left[ \textrm{OIII} \right]$.

SDSS1715+6008 (Fig. \ref{F:twod_1715}) shows the least complex 2D spectrum (Fig. \ref{F:twod_1715}). The gas in SDSS1715+6008 appears to be rotating in a disk extending out to several tens of kpc. The pattern of rotation is relatively smooth in SDSS1715+6008 (Fig. \ref{F:kinematics}). Both MgIb and NaID are detected, allowing the stellar kinematics to be traced. The stellar lines however show rotation in exactly the opposite direction. This pattern can either be interpreted as counter-rotation of gas with respect to stars, or as a sign of a bi-conical outflow.

\section{Discussion}
\label{S:discussion}

\subsection{Merger Stages}

\subsubsection{Early-stage mergers}

Of the four sources studied here, three were selected because they show double-nucleated structures in the NIR, with each nucleus associated with a separate [OIII] component. All objects have colors placing them in the red sequence, consistent with dust-reddened starbursts, as expected for such early merger stages. Larger reddening in the central region, co-spatial with compact u-band emission indicates compact star-formation in the central region. All of these sources show tidal features as well as complex NLRs, with multiple $\left[ \textrm{OIII} \right]$ clumps separated spatially or in velocity. Very extended ($\sim$50-100 kpc) narrow $\left[ \textrm{OIII} \right]$ features with little or no velocity gradients are seen in two objects (SDSS0952+2552, Fig. \ref{F:twod_0952} and SDSS1502+1115, Fig. \ref{F:twod_1502}), similar to features detected in luminous obscured quasars \citep{greene_feedback_2011}. In both objects, these features are blue-shifted with respect to the closest NLR feature, they could therefore be coasting gas shells ejected during the merging process. The overall NLRs are extremely extended, reaching sizes up to $\sim$100 kpc. Such NLR sizes are about one order of magnitude larger than those of typical AGN of comparable luminosity \citep{bennert_size_2002,bennert_size_2006,schmitt_hubble_2003,greene_feedback_2011} and are more consistent with NLRs in AGN about one order of magnitude brighter than those studied here \citep{humphrey_integral-field_2010,greene_feedback_2011,liu_observations_2013}. 

Close companions are apparent in all three sources that show double-nuclei in the NIR. In two cases, the companion are at the redshift of the AGN (see Figures \ref{F:lines_0952}, \ref{F:lines_1502}), in the other case, no emission or absorption lines are detected in the companion, making a redshift determination impossible. The companion of SDSS0952+2552 shows emission in $\left[ \textrm{OIII} \right]$ (Fig. \ref{F:lines_0952}), although this could also be due to projected extended NLR emission. In the spectra of the companion of SDSS1502+1115, we detect Balmer lines in absorption as well as MgIb (Fig. \ref{F:lines_1502}), indicative of a young stellar population \citep{bressan_probing_1996}. The companion of SDSS1502+1115 is unresolved, and therefore has a size of $\leqslant$ 5 kpc. This companion of SDSS1502+1115 could be a tidal dwarf created during the interaction \citep{bournaud_high-resolution_2008}, this scenario can explain its compactness, the absorption lines consistent with young stellar populations as well as its association with an ongoing merger. While the numbers are too small to draw clear conclusions, the high incidence ($>$2/3) of companions might either be a sign of double-peaked $\left[ \textrm{OIII} \right]$ emitters being preferentially located in dense environment. The velocity offsets between the AGN and companion ($\Delta$ v $\sim$ 100/600 km s$^{-1}$ for SDSS0952+2552/SDSS1502+1115 respectively, see Fig. \ref{F:kinematics}) are consistent with the velocity dispersion of low to intermediate mass galaxy groups \citep{balogh_galaxy_2004}. To test if the incidence of nearby neighbours is consistent with expected densities of galaxies from clustering, we estimate the probability of finding a nearby neighbour within a volume dV using the following function:

\begin{equation}
dP = \Phi \times (1+ \xi (r)) \times dV
\end{equation} 

where $\Phi$ is the number density of galaxies and $\xi(r)$ is the two-point correlation function. Using two point correlation functions from \citet{zehavi_luminosity_2005} and number densities from \citet{blanton_galaxy_2003}, the probability of finding a nearby neighbour at the given radius can be estimated. Using the two-point correlation function value for r=200 kpc, integrating the $r$ band galaxy luminosity function from \citep{blanton_galaxy_2003} in the range -23 $\leq M_r \leq$ -21 yields and integrating over a sphere where the radius is set to the distance of the neighbour, gives an expected number of neighbours $\geqslant$ 1. The incidence of nearby neighbours is therefore consistent with small-scale clustering.

The high rate of close companions found is also consistent with the findings of other work studying the properties of double-peaked $\left[ \textrm{OIII} \right]$ emitters. \citet{rosario_adaptive_2011} found that about 1/3 of double-peaked $\left[ \textrm{OIII} \right]$ AGN have nearby companions ($\leqslant$ 50 kpc), inconsistent with background object counts. 

\subsubsection{Undisturbed X-ray binary}

The X-ray binary AGN SDSS1715+6008 shows no signs of recent interaction in the imaging data. The colors place the object in the red sequence. The color map shows a wedge feature, consistent with higher dust extinction or supression of $\left[ \textrm{OIII} \right]$ emission in this region. No signs of compact star-formation are observed in this object. Spectroscopy shows de-coupling between the gas and stars, with both components showing signs of rotation. This can be explained as a sign of a bipolar outflow \citep[e.g.][]{fischer_hubble_2011} or a gas disk counter-rotating with respect to the stellar disk. Counter-rotating gas disks can be explained by accretion of a gas-rich galaxy onto a disk with relatively low gas content \citep[see e.g.][]{morse_inclined_1998,coccato_dating_2011,coccato_spectroscopic_2013}. Given the ubiquity of outflows in AGN, the outflow scenario is favoured over the counter-rotating gas-disk scenario. In any case, the binary AGN in SDSS1715+6008 does not appear to be the origin of the double-peaked $\left[ \textrm{OIII} \right]$ emission in this source. The blue and red $\left[ \textrm{OIII} \right]$ components are not each centred on one of the X-ray sources. The NLR rather shows velocity gradients across the galaxy consistent with either rotation or a bipolar outflow. This shows that bipolar outflows can show double-peaked $\left[ \textrm{OIII} \right]$ structure, as already noted in the literature \citep[e.g.][]{shen_binary_2010,smith_search_2010}.

The lack of merger features indicates that the event that created the binary black hole took place at least $\sim$ 0.5-1 Gyr ago \citep{lotz_effect_2010,lotz_effect_2010-1}. Such long time-scales are on the upper end of coalescence time-scales for binary black hole mergers from simulations \citep{khan_mergers_2012}, especially if the initial merger was gas-poor. These long time-scales are also consistent with studies showing indications of displaced or recently merged binary black holes, such as the finding that the central black hole in M87 is displaced by $\sim$ 7 pc \citep{batcheldor_displaced_2010}, indicative of a recoil from a merging event.

The X-ray binary AGN SDSS1715+6008 is a fascinating source. It is a double-peaked $\left[ \textrm{OIII} \right]$ emitter, but the binary AGN is not the source of the double-peaked $\left[ \textrm{OIII} \right]$. The low disturbance levels in the galaxy indicate that coalescence of black holes in mergers can take longer than it takes for merger features to dissipate. This suggests the existence of binary black hole systems in relaxed galaxies and might provide an interesting path to tracing past merger history of galaxies.

\subsection{High velocity outflows}

Of the three NIR double-nucleated AGN (SDSS0952+2552, SDSS1151+4711, SDSS1502+1115), all show broad predominantly blue wings in $\left[ \textrm{OIII} \right]$ (see Figures \ref{F:spec_0952}, \ref{F:spec_1151}, \ref{F:spec_1502}). These high-velocity outflows (v $\leqslant$ 1500 km $s^{-1}$) are only observed in the very central regions, co-spatial with the nucleus. HeII emission and in one case even $\left[ \textrm{FeVII} \right]$ (a coronal line associated with quasar outflows \citeauthor{mullaney_location_2009} \citeyear{mullaney_location_2009}; \citeauthor{ward_highly_2010} \citeyear{ward_highly_2010}) is observed co-spatially with the outflows (see Figures \ref{F:lines_1151}, \ref{F:lines_1502}), indicative of either photo-ionization by the AGN or strong shocks. The outflows observed in the ionized gas are likely driven or at least bolstered by the AGN since starburst driven winds have limiting speeds $\lesssim$1000 km s$^{-1}$ \citep{strickland_starburst-driven_2000,thacker_quasars:_2006,sharma_supernovae_2013}. However, observations indicate that higher speeds might be possible in some systems \citep{chung_evidence_2011}.

The highest velocity outflows are co-spatial with the central nucleus showing properties consistent with a starburst. Theoretical models of galaxy formation predict that fast outflows from AGN can quench star formation in galaxies \citep[e.g.][]{di_matteo_energy_2005,hopkins_cosmological_2008}. Signs of such negative feedback from AGN are observed in some sources \citep[e.g.][]{rupke_integral_2011,cano-diaz_observational_2012,farrah_direct_2012}. In our data, we see no signs of negative feedback. The high incidence of the high-velocity outflows observed in ionized gas in our sample compared to the general AGN population implies that either the AGN are young or that the co-existence with either the starburst or the ongoing merger increase the probability for fast outflows.

\subsection{Are the double-peaked $\left[ \textrm{OIII} \right]$ emitters binary AGN?}

For SDSS1715+6008, the identification as a binary AGN stems from X-ray data \citep{comerford_chandra_2011}, our data cannot assess this identification since the separation is well below the resolution of our data. However, the double-peaked $\left[ \textrm{OIII} \right]$ emission is consistent with a bi-conical outflow.

For the remaining three sources, the nuclei were resolved in our imaging and spectroscopic data. However, we do not find unequivocal evidence of AGN activity in the secondary core. While all three sources show narrow line emission indicative of ionization by AGN, illumination by the Type 1 AGN detected in all sources cannot be excluded. Broad emission lines in the secondary core would be clear evidence of AGN activity in the second core since the BLR is located $<<$ 1 pc rather than kpcs from the black hole. The line ratios are not found to vary across the slit. Observations of such variations could indicate differences in the shape of the ionizing continuum and therefore ionization by two separate sources.

To conclude, the deep imaging and long-slit spectroscopy did not provide enough information to identify any of the sources as binary AGN. Due to the large sizes of the NLRs in the objects observed, an emission line spectrum with line ratios indicative of AGN detected in the secondary core is not sufficient for identifying AGN since illumination by a single AGN cannot be excluded. Detection of broad-line emission in both cores would be a clear sign of a binary AGN. However, this was not detected in our sample. Lack of detection of such features can either indicate the remaining objects are not binary AGN or that the broad-line emitting region in the secondary core is shielded by toroidal obscuration. X-ray and radio observations are likely more promising tools for confirming binary AGN since they are less affected by toroidal obscuration. Indeed, two of the sources in our sample have been confirmed using those methods, despite not showing clear signs of being binary AGN in the optical data.

\section{Conclusions}
\label{S:conclusions}

We present deep $u,r,z$ imaging and medium resolution long-slit spectroscopy of four double-peaked $\left[ \textrm{OIII} \right]$ AGN. Double-peaked $\left[ \textrm{OIII} \right]$ emitting AGN form a complex class consisting of binary AGN, major mergers hosting a single AGN and AGN with complex NLR kinematics. We selected a sample of four double-peaked $\left[ \textrm{OIII} \right]$ AGN, two of which are confirmed binary AGN, the remaining two are candidate binary AGN located in major mergers with at least one AGN. The four objects studied here are radio-quiet to radio-intermediate, luminous AGN ($L_{bol} \sim 10^{44}$ erg/s).

\begin{itemize}
\item Three AGN selected for their double-nucleated structure in the NIR show clear signs of ongoing merging with unambiguous tidal features. The NLRs have complex kinematics and sizes of up to 100 kpc, comparable to the most extended NLRs observed in luminous quasars \citep[e.g.][]{greene_feedback_2011, liu_observations_2013}. The NLR also contains compact clumps of sizes $\leqslant$ 5 kpc, consistent with NLR sizes for AGN with similar luminosities \citep{bennert_size_2006,greene_feedback_2011}. The galaxies have redder colors in the nuclear regions, consistent with central starbursts. All three NIR selected AGN have nearby neighbours, in two cases, these are at the same redshift as the AGN. The high incidence of close companions is consistent with expected numbers from small-scale clustering properties.
\item Powerful and fast outflows ($\sim$ 1500 km s$^{-1}$) in the ionized gas are observed in all three AGN currently undergoing merging. The outflows do not extend throughout the galaxy but are compact and co-spatial with the central starbursts. The large outflow speeds indicate that the outflows are driven by the AGN. Suppression of star formation co-spatial with the outflows is not observed.  The high incidence of powerful outflows in our sample suggests that either the presumed youth of the AGN, the association with mergers or a connection with ongoing starbursts increases the probability of fast outflows.
\item The X-ray binary SDSS1715+6008 \citep{comerford_1.75_2009} shows no signs of recent interaction down to low surface brightness levels ($\sim$ 27 mag/arcsec$^{2}$ in $r$). This suggests that the event that formed the binary black hole lies at least $\sim 1 Gyr$ in the past. The ionized gas and stellar kinematics are misaligned, consistent with either bipolar outflows or a counter-rotating gas disk. The detection of a binary AGN in a relaxed host galaxy implies that coalescence of binary black holes can take as long as $\sim$ 1 Gyr. Binary black holes and AGN could therefore be present in a number of dynamically relaxed galaxies. 
\item Using the deep imaging data and long-slit spectroscopy presented here, we could not clearly identify the objects studied as binary AGN. Optical data alone is not always sufficient to identify binary AGN since obscuration by the torus strongly affects unambiguous signs of AGN activity. X-ray and radio data are more powerful methods for identifying binary AGN since they are less affected by obscuration. 
\end{itemize}

Whilst the sample studied here is too small to draw statistical conclusions, we have presented an in-depth study of the physical properties of doubled-peaked $\left[ \textrm{OIII} \right]$ emitters. In particular the large incidence of fast outflows in ionized gas and nearby companions warrant further study. The discovery of a binary AGN located in an undisturbed host galaxy gives important observational hints for the time-scales of black hole coalescence in mergers.

\section*{Acknowledgements}
We thank the referee for his/her comments and suggestions. This work was supported in part by the USA National Science Foundation grant AST-1009628. Based on observations made with the Gran Telescopio Canarias (GTC), installed in the Spanish Observatorio del Roque de los Muchachos of the Instituto de Astrofísica de Canarias, in the island of La Palma. We would like to thank Mike Crenshaw, David Rosario, Carole Mundell and Rita Tojeiro for discussion and comments.

\bibliographystyle{apj}
\bibliography{gtc}

\appendix

\section{Line Fits}
\label{S:appendix}

All lines are fit using \textsc{PySpecKit}\footnote{http://pyspeckit.bitbucket.org/html/sphinx/index.html}. Line widths are given as FWHM converted to units of km s$^{-1}$. In most cases, each emission line is fit separately. Lines are only fit jointly when problems arise in the separate line fits due to merging of lines or close nearby lines. In particular, joint line fits are performed in the following cases:

\begin{itemize}
\item For SDSS0952+2552, H$\gamma$ and $\left[ \textrm{OIII} \right]$ 4363 are fit jointly but leaving both velocity or line widths unconstrained for both lines.
\item For SDSS1151+4711, H$\gamma$ and $\left[ \textrm{OIII} \right]$ 4363 are fit jointly but leaving both velocity or line widths unconstrained for both lines.
\item For SDSS1502+1115, the MgIb triplet is fit as three lines locked in velocity as well as line widths, line ratios are fixed to atomic values.
\item For SDSS1715+6008, there is a a strong blend between both [NI] 5200 and MgIb as well as NaID and HeI5876. To achieve best fits, those 4 features are fit jointly as follows: NI and HeI are locked in both velocity and line widths. MgIb is fit as three lines locked in velocity as well as line widths, line ratios are fixed to atomic values. NaID accordingly is fit with lines locked in velocity and line widths to MgIb, as with MgIb, line ratios are fixed at atomic values, but line ratios between NaID and MgIb are left to vary. This results in a joint fit for a gas (NI+HeI) and stellar (MgIb+NaID) component.  
\item For all objects, the fits for H$\beta$ are inspected and refit with both a broad and narrow component if necessary.
\item For all objects, if no stable fit can be found in a given slice, the line is treated as a non-detection in the given slice and no values are given in the tables.
\end{itemize}

\clearpage

\begin{figure*}
\begin{center}
\includegraphics[width=16cm]{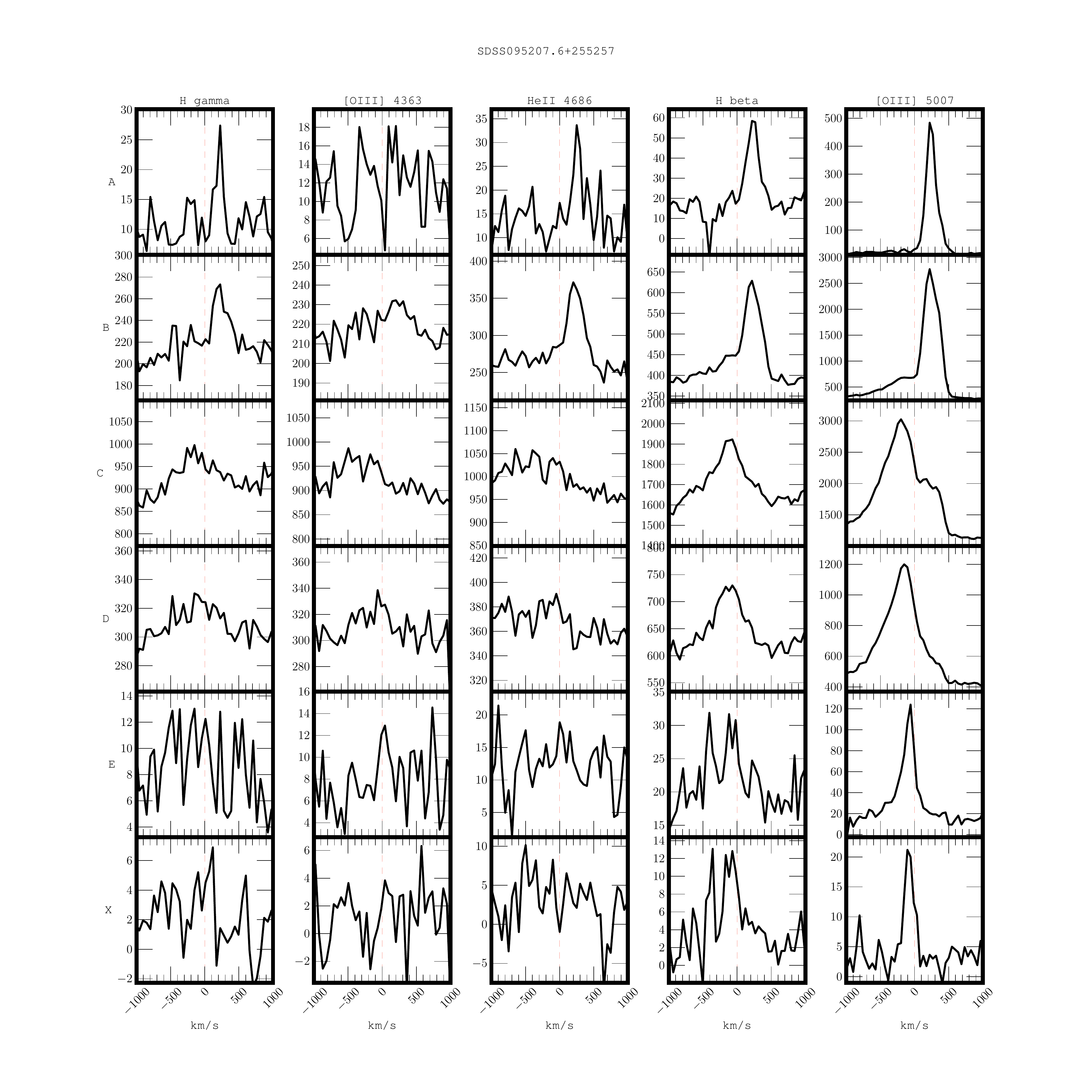}
\caption{Changes of shapes in notable lines of SDSS0952+2552 across the slit. The location of the extraction boxes are shown in Fig. \ref{F:slitmaps}, upper left panel. The top panel corresponds to the most southern extraction box, and the lowest panel to the most northern extraction box. Arbitrary flux units are shown for each panel to show relative line strengths.
\label{F:lines_0952}}
\end{center}
\end{figure*}

\begin{figure*}
\begin{center}
\includegraphics[width=20cm]{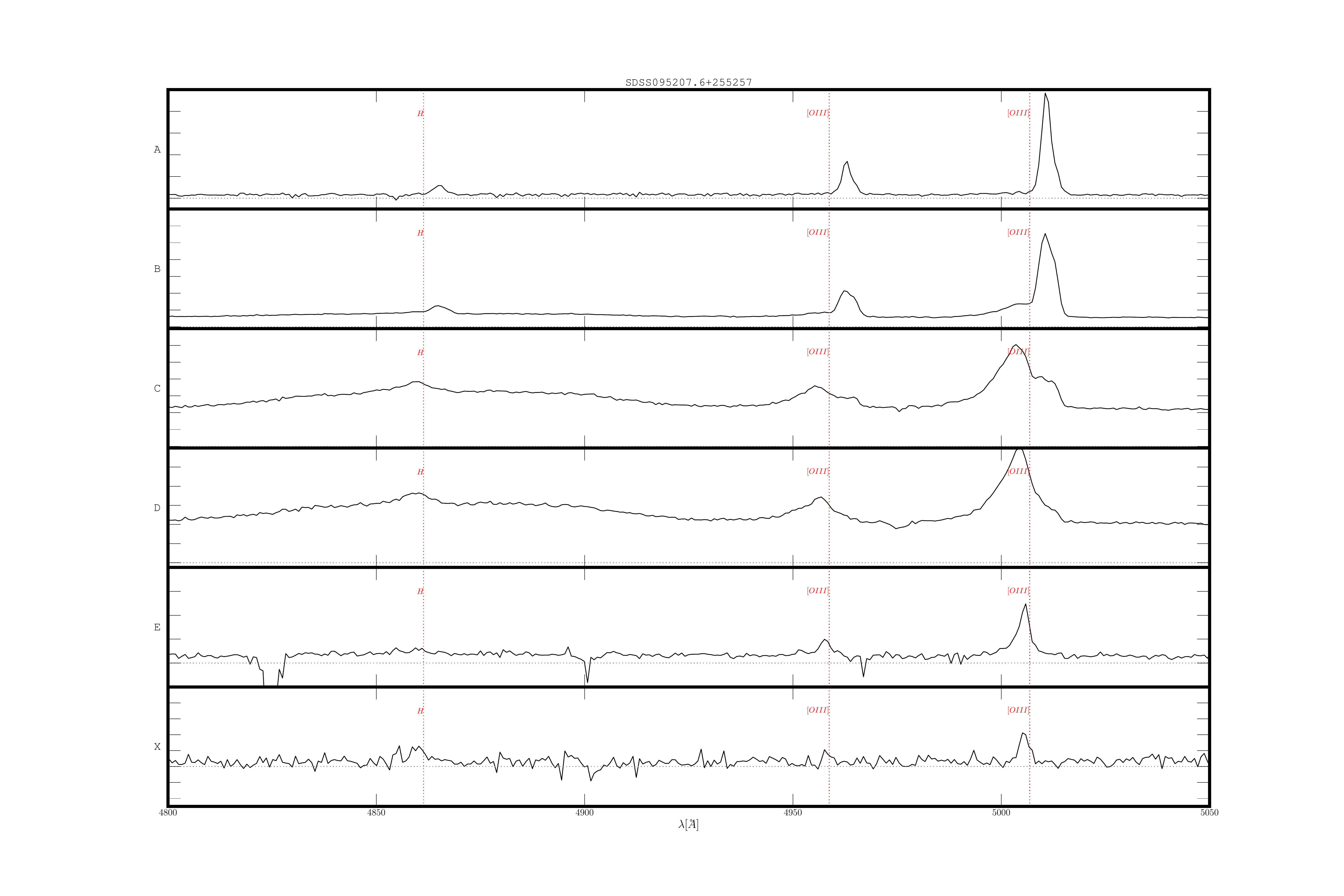}
\caption{1D spectra extracted along the slit for SDSS0952+2552 showing H$\beta$ and $\left[ \textrm{OIII} \right]$. The location of the extraction boxes are shown in Fig. \ref{F:slitmaps}, upper left panel. The top panel corresponds to the most southern extraction box, and the lowest panel to the most northern extraction box. Each panel is scaled to best show the spectral features, refer to Fig. \ref{F:lines_0952} for detailed shapes of individual lines. A dotted horizontal line shows zero flux in all panels. If not visible, the axis is at the same position as the dotted line for a given panel.
\label{F:spec_0952}}
\end{center}
\end{figure*}

\begin{figure*}
\begin{center}
\includegraphics[width=16cm]{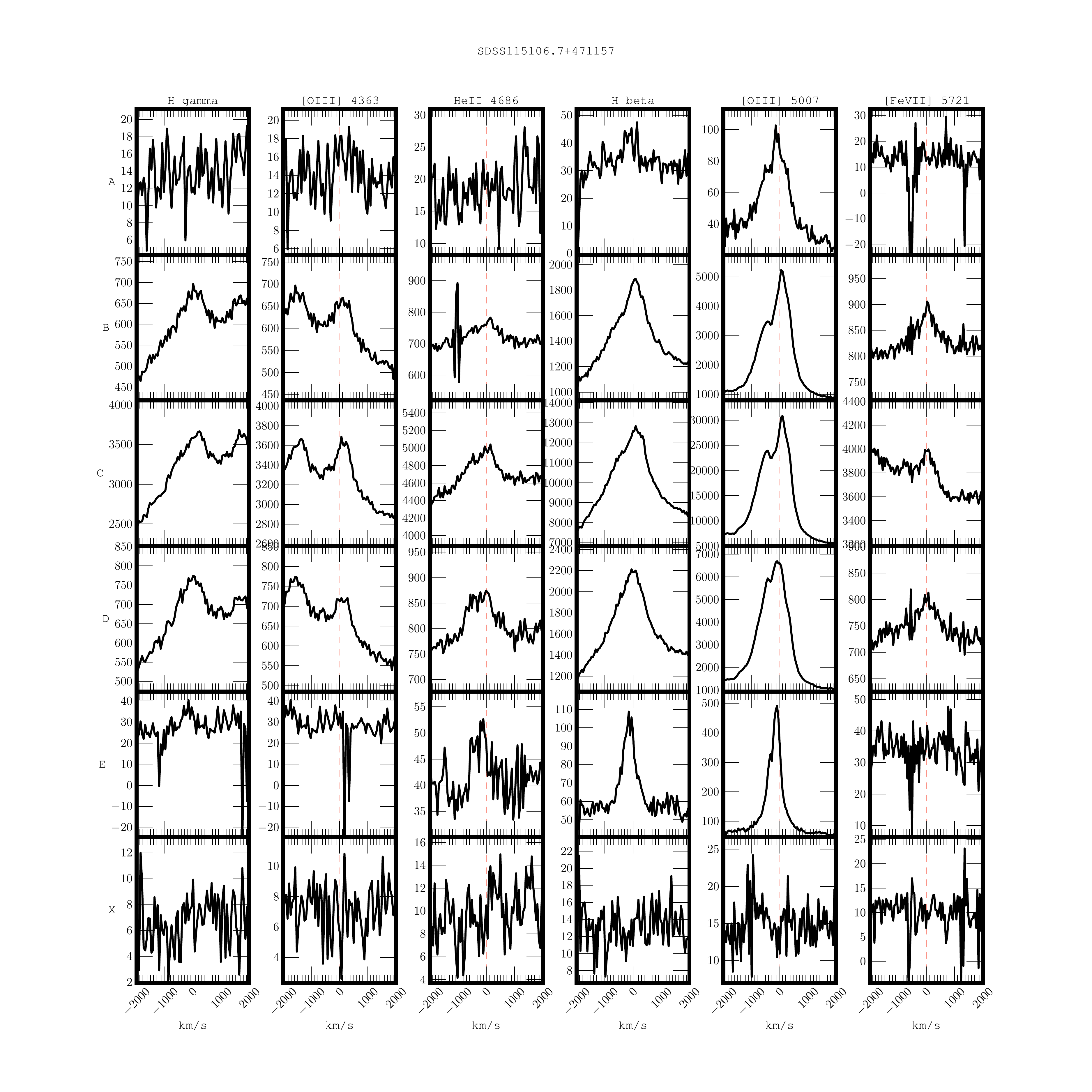}
\caption{Changes of shapes in notable lines of SDSS1151+4711 across the slit. The location of the extraction boxes are shown in Fig. \ref{F:slitmaps}, upper right panel. The top panel corresponds to the most southern extraction box, and the lowest panel to the most northern extraction box. Arbitrary flux units are shown for each panel to show relative line strengths.
\label{F:lines_1151}}
\end{center}
\end{figure*}

\begin{figure*}
\begin{center}
\includegraphics[width=20cm]{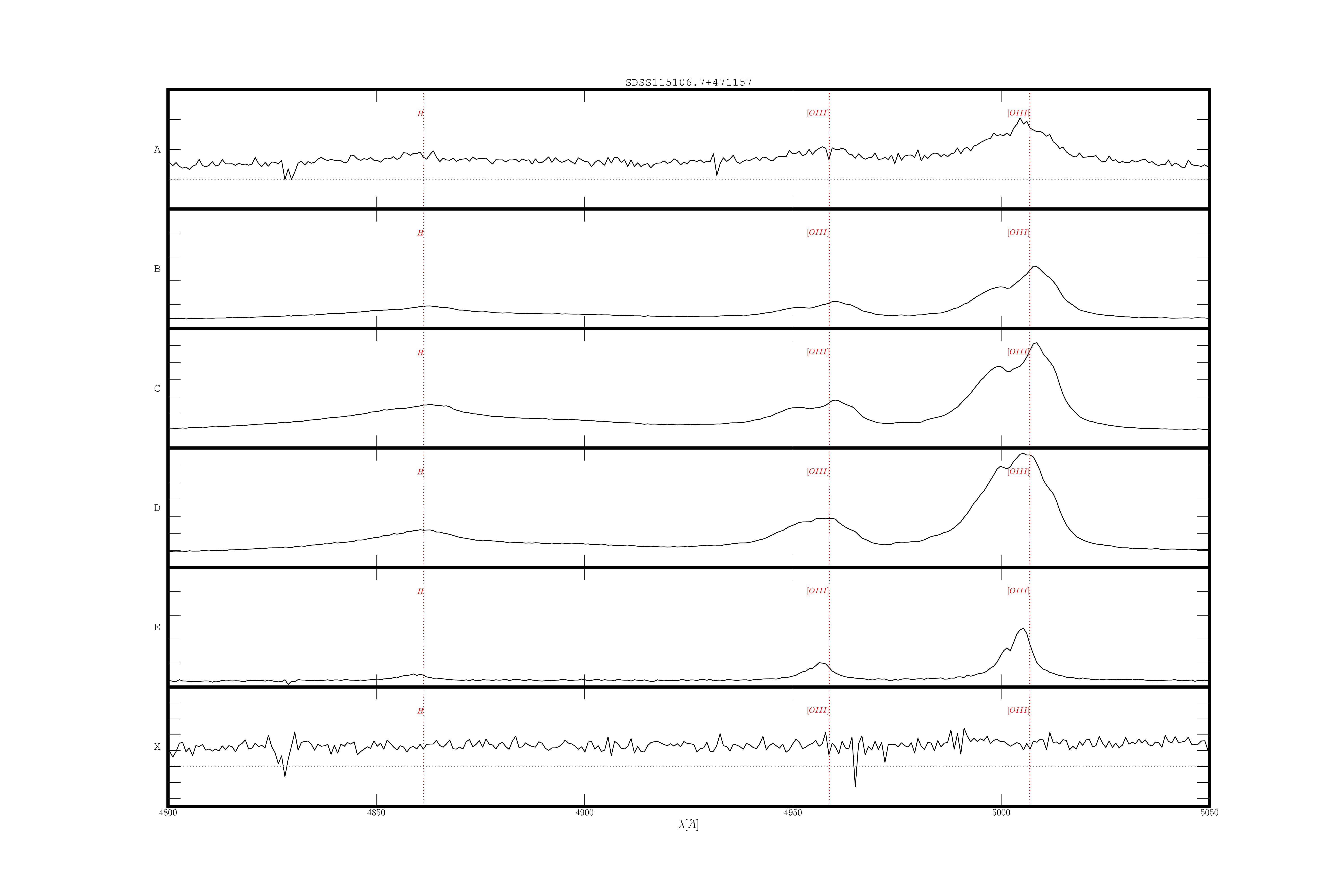}
\caption{1D spectra extracted along the slit for SDSS1151+4711 showing H$\beta$ and $\left[ \textrm{OIII} \right]$. The location of the extraction boxes are shown in Fig. \ref{F:slitmaps}, upper right panel. The top panel corresponds to the most southern extraction box, and the lowest panel to the most northern extraction box. Each panel is scaled to best show the spectral features, refer to Fig. \ref{F:lines_1151} for detailed shapes of individual lines. A dotted horizontal line shows zero flux in all panels. If not visible, the axis is at the same position as the dotted line for a given panel.
\label{F:spec_1151}}
\end{center}
\end{figure*}

\begin{figure*}
\begin{center}
\includegraphics[width=16cm]{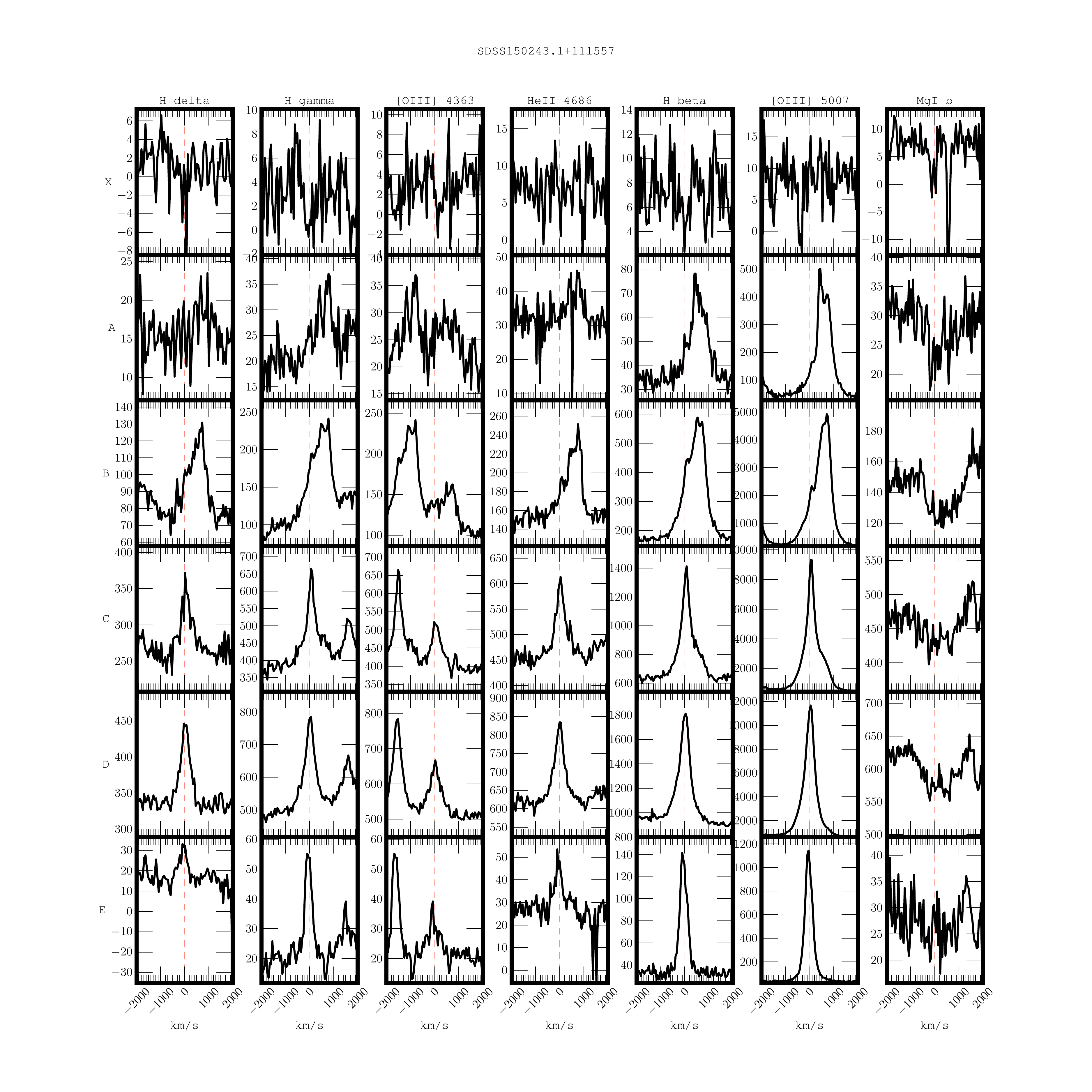}
\caption{Changes of shapes in notable lines of SDSS1502+1115 across the slit. The location of the extraction boxes are shown in Fig. \ref{F:slitmaps}, lower left panel. The top panel corresponds to the most southern extraction box, and the lowest panel to the most northern extraction box. Arbitrary flux units are shown for each panel to show relative line strengths.
\label{F:lines_1502}}
\end{center}
\end{figure*}

\begin{figure*}
\begin{center}
\includegraphics[width=20cm]{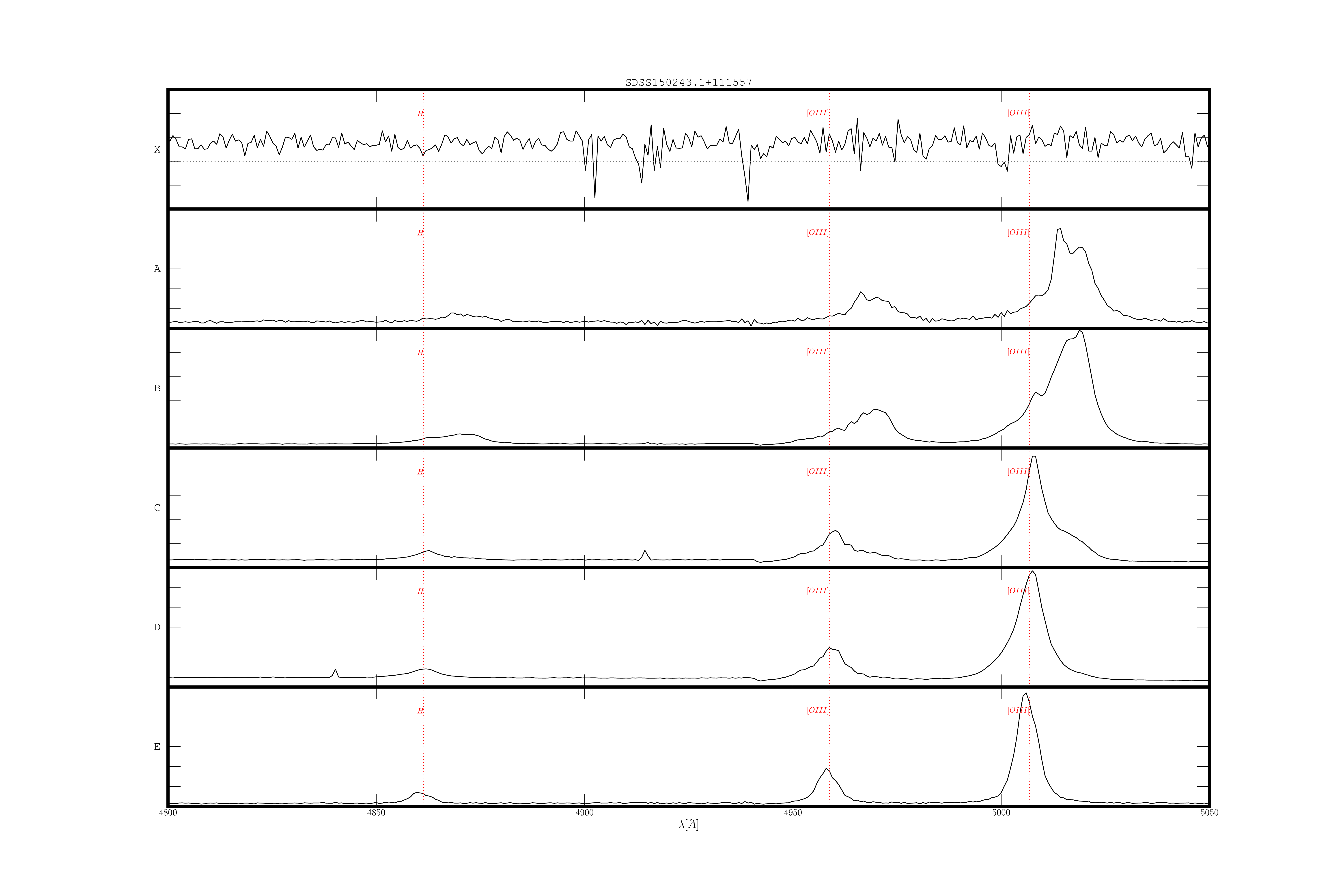}
\caption{1D spectra extracted along the slit for SDSS1502+1115 showing H$\beta$ and $\left[ \textrm{OIII} \right]$. The location of the extraction boxes are shown in Fig. \ref{F:slitmaps}, lower left panel. The top panel corresponds to the most southern extraction box, and the lowest panel to the most northern extraction box. Each panel is scaled to best show the spectral features, refer to Fig. \ref{F:lines_1502} for detailed shapes of individual lines. A dotted horizontal line shows zero flux in all panels. If not visible, the axis is at the same position as the dotted line for a given panel.
\label{F:spec_1502}}
\end{center}
\end{figure*}

\begin{figure*}
\begin{center}
\includegraphics[width=16cm]{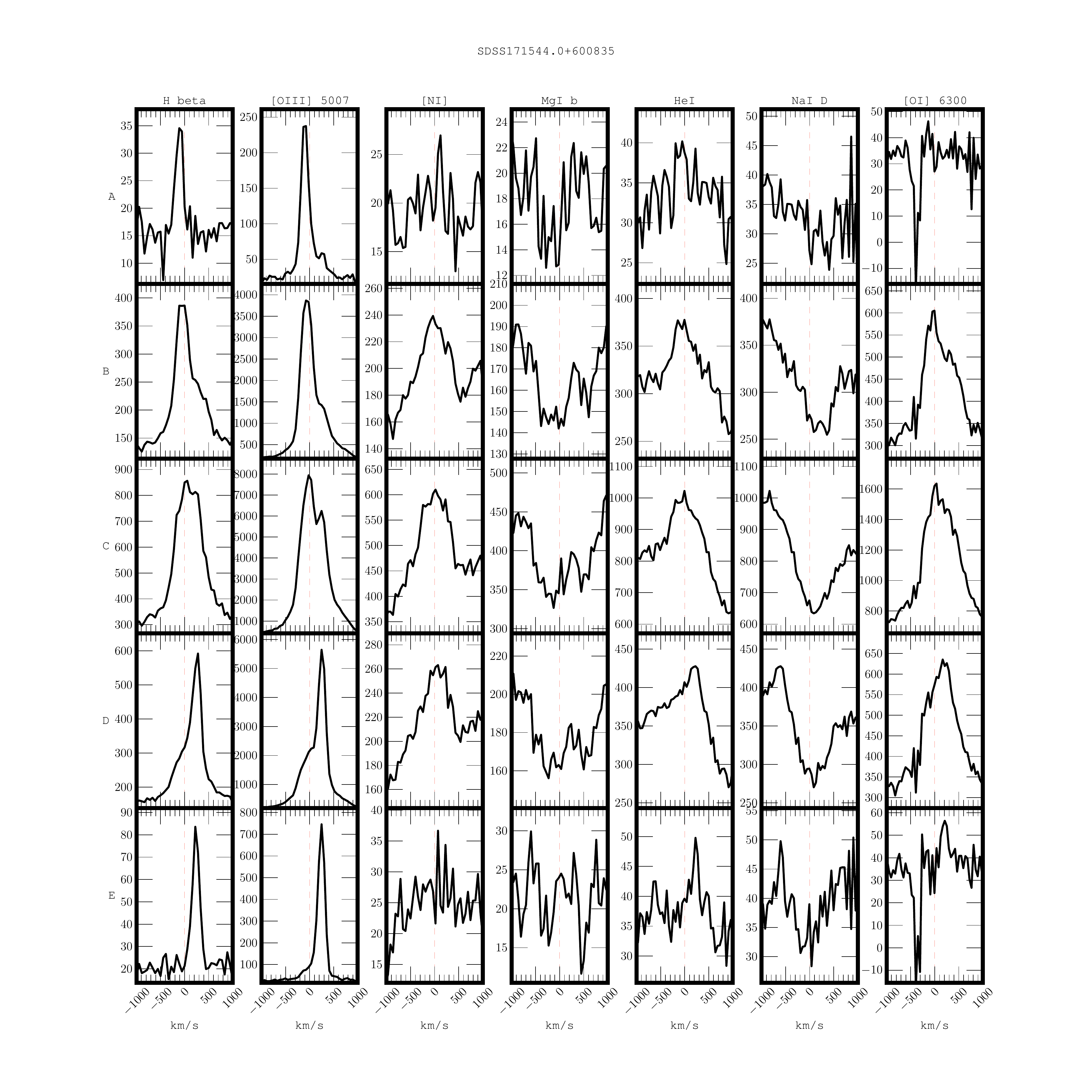}
\caption{Changes of shapes in notable lines of SDSS1715+6008 across the slit. The location of the extraction boxes are shown in Fig. \ref{F:slitmaps}, lower right panel. The top panel corresponds to the most southern extraction box, and the lowest panel to the most northern extraction box. Arbitrary flux units are shown for each panel to show relative line strengths.
\label{F:lines_1715}}
\end{center}
\end{figure*}

\begin{figure*}
\begin{center}
\includegraphics[width=20cm]{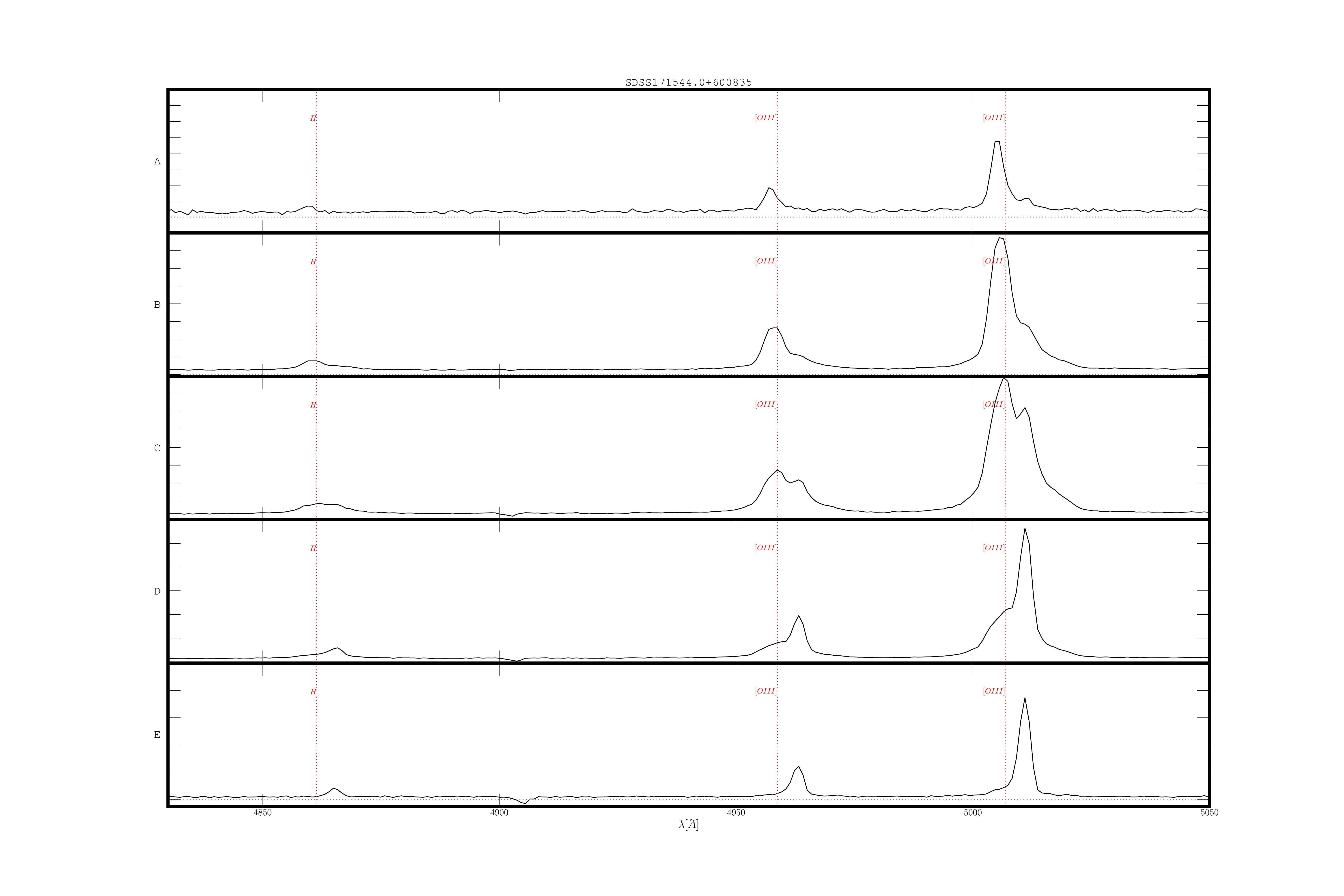}
\caption{1D spectra extracted along the slit for SDSS1715+6008 showing H$\beta$ and $\left[ \textrm{OIII} \right]$. The location of the extraction boxes are shown in Fig. \ref{F:slitmaps}, lower right panel. The top panel corresponds to the most southern extraction box, and the lowest panel to the most northern extraction box. Each panel is scaled to best show the spectral features, refer to Fig. \ref{F:lines_1715} for detailed shapes of individual lines. A dotted horizontal line shows zero flux in all panels. If not visible, the axis is at the same position as the dotted line for a given panel.
\label{F:spec_1715}}
\end{center}
\end{figure*}

\begin{table}
\caption{Spatially resolved kinematics of lines for SDSS0952+2552. Continued in Table \ref{T:lines_0952b}. The lines are fitted in five spatial bins across the slit, as well as one additional bin for objects in which either tidal tails or a nearby neighbour falls on the slits. The location of those spatial bins are indicated in Fig. \ref{F:slitmaps} and are identical to those shown in Figures \ref{F:lines_0952} and \ref{F:spec_0952}. All velocities are given with respect to the nominal object redshift listed in Table \ref{T:objects}. Negative velocities indicate a blueshift and positive velocities a redshift.}
\begin{tabular}{ccrrrr}
\hline
Line & Slice & v & $\Delta$ v & $FWHM$ & $\Delta FWHM$\\
         &          &  [km s$^{-1}$] &  [km s$^{-1}$]  & [km s$^{-1}$] &  [km s$^{-1}$] \\
\hline
$H_{\gamma}$ & A & -- & -- & -- & -- \\ 
$H_{\gamma}$ & B & 210.75 & 18.83 & 517.37 & 45.01 \\ 
$H_{\gamma}$ & C & -229.59 & 49.42 & 1026.92 & 100.71 \\ 
$H_{\gamma}$ & D & -- & -- & -- & -- \\ 
$H_{\gamma}$ & E & -- & -- & -- & -- \\ 
$H_{\gamma}$ & X & -- & -- & -- & -- \\ 
$\left[ \textrm{OIII} \right]$ 4363 & A & -- & -- & -- & -- \\ 
$\left[ \textrm{OIII} \right]$ 4363 & B & 74.09 & 59.64 & 1183.6 & 155.74 \\ 
$\left[ \textrm{OIII} \right]$ 4363 & C & -248.72 & 72.2 & 1658.36 & 181.46 \\ 
$\left[ \textrm{OIII} \right]$ 4363 & D & -- & -- & -- & -- \\ 
$\left[ \textrm{OIII} \right]$ 4363 & E & -- & -- & -- & -- \\ 
$\left[ \textrm{OIII} \right]$ 4363 & X & -- & -- & -- & -- \\ 
HeII 4686 & A & 258.0 & 11.13 & 145.5 & 26.09 \\ 
HeII 4686 & B & 221.73 & 6.87 & 286.27 & 15.95 \\ 
HeII 4686 & C & -67.75 & 29.77 & 163.86 & 69.65 \\ 
HeII 4686 & D & -124.36 & 42.42 & 270.77 & 99.45 \\ 
HeII 4686 & E & -- & -- & -- & -- \\ 
HeII 4686 & X & -- & -- & -- & -- \\ 
\hline
\end{tabular}
\label{T:lines_0952}
\end{table}

\begin{table}
\caption{Table \ref{T:lines_0952} continued.}
\begin{tabular}{ccrrrr}
\hline
Line & Slice & v & $\Delta$ v & $FWHM$ & $\Delta FWHM$\\
         &          &  [km s$^{-1}$] &  [km s$^{-1}$]  & [km s$^{-1}$] &  [km s$^{-1}$] \\
\hline
$H_{\beta}$ & A & 229.52 & 6.75 & 210.19 & 15.7 \\ 
$H_{\beta}$ & B & 185.21 & 5.6 & 569.37 & 13.12 \\ 
$H_{\beta}$ & C & -153.41 & 13.69 & 585.14 & 40.29 \\ 
$H_{\beta}$ & D & -130.76 & 12.03 & 439.82 & 33.49 \\ 
$H_{\beta}$ & E & -188.74 & 64.81 & 927.56 & 152.44 \\ 
$H_{\beta}$ & X & -111.35 & 40.43 & 448.06 & 95.29 \\ 
$\left[ \textrm{OIII} \right]$ 4959 & A & 245.13 & 2.04 & 180.57 & 4.67 \\ 
$\left[ \textrm{OIII} \right]$ 4959 & B & 243.98 & 1.24 & 270.67 & 2.7 \\ 
$\left[ \textrm{OIII} \right]$ 4959 & C & -169.15 & 7.09 & 458.29 & 16.51 \\ 
$\left[ \textrm{OIII} \right]$ 4959 & D & -172.58 & 5.94 & 407.0 & 13.84 \\ 
$\left[ \textrm{OIII} \right]$ 4959 & E & -73.78 & 8.4 & 195.31 & 19.62 \\ 
$\left[ \textrm{OIII} \right]$ 4959 & X & -61.8 & 16.6 & 87.39 & 43.86 \\ 
$\left[ \textrm{OIII} \right]$ 5007 & A & 244.15 & 0.82 & 187.21 & 1.82 \\ 
$\left[ \textrm{OIII} \right]$ 5007 & B & 239.3 & 0.53 & 283.85 & 1.09 \\ 
$\left[ \textrm{OIII} \right]$ 5007 & C & -180.43 & 2.52 & 781.62 & 5.7 \\ 
$\left[ \textrm{OIII} \right]$ 5007 & D & -192.78 & 2.2 & 625.46 & 5.01 \\ 
$\left[ \textrm{OIII} \right]$ 5007 & E & -87.98 & 3.41 & 241.48 & 7.91 \\ 
$\left[ \textrm{OIII} \right]$ 5007 & X & -80.85 & 9.52 & 153.61 & 22.27 \\ 
\hline
\end{tabular}
\label{T:lines_0952b}
\end{table}

\begin{table}
\caption{Spatially resolved kinematics of lines for SDSS1151+4711. Continued in Table \ref{T:lines_1151b}. The lines are fitted in five spatial bins across the slit, as well as one additional bin for objects in which either tidal tails or a nearby neighbour falls on the slits. The location of those spatial bins are indicated in Fig. \ref{F:slitmaps} and are identical to those shown in Figures \ref{F:lines_0952} and \ref{F:spec_0952}. All velocities are given with respect to the nominal object redshift listed in Table \ref{T:objects}. Negative velocities indicate a blueshift and positive velocities a redshift.}
\begin{tabular}{ccrrrr}
\hline
Line & Slice & v & $\Delta$ v & $FWHM$ & $\Delta FWHM$ \\
      &          &  [km s$^{-1}$] &  [km s$^{-1}$]  & [km s$^{-1}$] &  [km s$^{-1}$]  \\
\hline
$H_{\gamma}$ & A & -- & -- & -- & -- \\ 
$H_{\gamma}$ & B & 22.83 & 22.46 & 1621.87 & 57.24 \\ 
$H_{\gamma}$ & C & 94.97 & 13.23 & 1398.71 & 33.37 \\ 
$H_{\gamma}$ & D & -2.72 & 18.08 & 1605.3 & 46.36 \\ 
$H_{\gamma}$ & E & -- & -- & -- & -- \\ 
$H_{\gamma}$ & X & -- & -- & -- & -- \\ 
$\left[ \textrm{OIII} \right]$ 4363 & A & -- & -- & -- & -- \\ 
$\left[ \textrm{OIII} \right]$ 4363 & B & 155.51 & 20.02 & 1061.94 & 45.74 \\ 
$\left[ \textrm{OIII} \right]$ 4363 & C & 130.08 & 11.97 & 962.51 & 27.97 \\ 
$\left[ \textrm{OIII} \right]$ 4363 & D & 133.43 & 20.49 & 1064.16 & 46.68 \\ 
$\left[ \textrm{OIII} \right]$ 4363 & E & -- & -- & -- & -- \\ 
$\left[ \textrm{OIII} \right]$ 4363 & X & -- & -- & -- & -- \\ 
HeII 4686 & A & -114.12 & 35.79 & 67.68 & 60.61 \\ 
HeII 4686 & B & -- & -- & -- & -- \\ 
HeII 4686 & C & -118.75 & 21.98 & 953.19 & 51.43 \\ 
HeII 4686 & D & -215.26 & 24.54 & 990.51 & 57.29 \\ 
HeII 4686 & E & -215.24 & 47.09 & 535.82 & 110.33 \\ 
HeII 4686 & X & -196.5 & 99.65 & 307.82 & 235.44 \\ 
\hline
\end{tabular}
\label{T:lines_1151}
\end{table}

\begin{table}
\caption{Table \ref{T:lines_1151} continued.}
\begin{tabular}{ccrrrr}
\hline
Line & Slice & v & $\Delta$ v & $FWHM$ & $\Delta FWHM$ \\
      &          &  [km s$^{-1}$] &  [km s$^{-1}$]  & [km s$^{-1}$] &  [km s$^{-1}$]  \\
\hline
$H_{\beta}$ & A & -239.12 & 68.55 & 1313.17 & 161.48 \\ 
$H_{\beta}$ & B & -57.38 & 4.7 & 1554.83 & 10.83 \\ 
$H_{\beta}$ & C & -125.55 & 2.78 & 1674.46 & 6.37 \\ 
$H_{\beta}$ & D & -121.89 & 3.92 & 1417.24 & 9.0 \\ 
$H_{\beta}$ & E & -146.57 & 11.07 & 568.56 & 25.81 \\ 
$H_{\beta}$ & X & -- & -- & -- & -- \\ 
$\left[ \textrm{OIII} \right]$ 4959 & A & -78.55 & 38.75 & 750.62 & 90.66 \\ 
$\left[ \textrm{OIII} \right]$ 4959 & B & -5.29 & 2.64 & 805.18 & 6.01 \\ 
$\left[ \textrm{OIII} \right]$ 4959 & C & -74.47 & 1.99 & 855.39 & 4.44 \\ 
$\left[ \textrm{OIII} \right]$ 4959 & D & -165.92 & 1.93 & 870.08 & 4.29 \\ 
$\left[ \textrm{OIII} \right]$ 4959 & E & -157.57 & 4.11 & 480.54 & 9.47 \\ 
$\left[ \textrm{OIII} \right]$ 4959 & X & -- & -- & -- & -- \\ 
$\left[ \textrm{OIII} \right]$ 5007 & A & -158.99 & 12.11 & 1073.61 & 28.04 \\ 
$\left[ \textrm{OIII} \right]$ 5007 & B & -94.16 & 1.04 & 1110.22 & 2.2 \\ 
$\left[ \textrm{OIII} \right]$ 5007 & C & -165.69 & 0.69 & 1149.08 & 1.43 \\ 
$\left[ \textrm{OIII} \right]$ 5007 & D & -213.43 & 0.78 & 1056.27 & 1.63 \\ 
$\left[ \textrm{OIII} \right]$ 5007 & E & -165.42 & 1.78 & 554.48 & 3.98 \\ 
$\left[ \textrm{OIII} \right]$ 5007 & X & -- & -- & -- & -- \\ 
$\left[ \textrm{FeVIIF} \right]$ & A & 47.35 & -- & -- & -- \\ 
$\left[ \textrm{FeVIIF} \right]$ & B & 89.19 & 27.62 & 1260.77 & 64.95 \\ 
$\left[ \textrm{FeVIIF} \right]$ & C & 58.49 & 20.14 & 346.35 & 47.3\\ 
$\left[ \textrm{FeVIIF} \right]$ & D & 75.97 & 30.36 & 1250.54 & 71.38\\ 
$\left[ \textrm{FeVIIF} \right]$ & E & 32.65 & 18.71 & 59.58 & 29.31\\ 
$\left[ \textrm{FeVIIF} \right]$ & X & 34.69 & -- & -- & -- \\ 
\hline
\end{tabular}
\label{T:lines_1151b}
\end{table}

\begin{table}
\caption{Spatially resolved kinematics of lines for SDSS1502+1115. Continued in Table \ref{T:lines_1502b}. The lines are fitted in five spatial bins across the slit, as well as one additional bin for objects in which either tidal tails or a nearby neighbour falls on the slits. The location of those spatial bins are indicated in Fig. \ref{F:slitmaps} and are identical to those shown in Figures \ref{F:lines_0952} and \ref{F:spec_0952}. All velocities are given with respect to the nominal object redshift listed in Table \ref{T:objects}. Negative velocities indicate a blueshift and positive velocities a redshift.}
\begin{tabular}{ccrrrr}
\hline
Line & Slice & v & $\Delta$ v & $FWHM$ & $\Delta FWHM$ \\
         &          &  [km s$^{-1}$] &  [km s$^{-1}$]  & [km s$^{-1}$] &  [km s$^{-1}$]  \\
\hline
$H_{\delta}$ & A & -- & -- & -- & -- \\ 
$H_{\delta}$ & B & 494.52 & 10.41 & 1123.72 & 23.99 \\ 
$H_{\delta}$ & C & 98.18 & 11.45 & 713.96 & 26.96 \\ 
$H_{\delta}$ & D & 13.33 & 5.44 & 550.59 & 12.65 \\ 
$H_{\delta}$ & E & -48.58 & 10.2 & 327.72 & 23.68 \\ 
$H_{\delta}$ & X & -17.01 & 100.06 & 314.78 & 235.84 \\ 
$H_{\gamma}$ & A & -- & -- & -- & -- \\ 
$H_{\gamma}$ & B & 483.61 & 10.03 & 1051.7 & 23.95 \\ 
$H_{\gamma}$ & C & 97.27 & 11.35 & 707.64 & 26.75 \\ 
$H_{\gamma}$ & D & 13.31 & 5.44 & 550.37 & 12.64 \\ 
$H_{\gamma}$ & E & -48.58 & 10.2 & 327.72 & 23.68 \\ 
$H_{\gamma}$ & X & -- & -- & -- & -- \\ 
$\left[ \textrm{OIII} \right]$ 4363 & A & -- & -- & -- & -- \\ 
$\left[ \textrm{OIII} \right]$ 4363 & B & 529.24 & 22.3 & 740.98 & 53.2 \\ 
$\left[ \textrm{OIII} \right]$ 4363 & C & 71.21 & 16.85 & 567.42 & 39.43 \\ 
$\left[ \textrm{OIII} \right]$ 4363 & D & 19.92 & 10.81 & 590.16 & 25.28 \\ 
$\left[ \textrm{OIII} \right]$ 4363 & E & -53.83 & 39.02 & 537.48 & 91.81 \\ 
$\left[ \textrm{OIII} \right]$ 4363 & X & -- & -- & -- & -- \\ 
\hline
\end{tabular}
\label{T:lines_1502}
\end{table}

\begin{table}
\caption{Table \ref{T:lines_1502} continued.}
\begin{tabular}{ccrrrr}
\hline
Line & Slice & v & $\Delta$ v & $FWHM$ & $\Delta FWHM$ \\
         &          &  [km s$^{-1}$] &  [km s$^{-1}$]  & [km s$^{-1}$] &  [km s$^{-1}$]  \\
\hline
HeII 4686 & A & -- & -- & -- & -- \\ 
HeII 4686 & B & 591.24 & 12.79 & 712.3 & 29.55 \\ 
HeII 4686 & C & 35.27 & 13.81 & 490.37 & 32.28 \\ 
HeII 4686 & D & -14.09 & 6.75 & 475.29 & 15.73 \\ 
HeII 4686 & E & -54.84 & 18.22 & 362.33 & 42.71 \\ 
HeII 4686 & X & -- & -- & -- & -- \\ 
$H_{\beta}$ & A & 559.78 & 15.51 & 793.81 & 36.15 \\ 
$H_{\beta}$ & B & 465.98 & 3.53 & 941.29 & 7.81 \\ 
$H_{\beta}$ & C & 81.79 & 4.42 & 654.63 & 10.29 \\ 
$H_{\beta}$ & D & -7.19 & 2.32 & 609.23 & 5.35 \\ 
$H_{\beta}$ & E & -49.02 & 3.93 & 370.02 & 8.98 \\ 
$H_{\beta}$ & X & 19.4 & 68.86 & 274.4 & 162.09 \\ 
$\left[ \textrm{OIII} \right]$ 4959 & A & 600.53 & 4.82 & 656.3 & 10.92 \\ 
$\left[ \textrm{OIII} \right]$ 4959 & B & 556.68 & 1.36 & 747.05 & 2.91 \\ 
$\left[ \textrm{OIII} \right]$ 4959 & C & 91.78 & 1.81 & 688.33 & 4.13 \\ 
$\left[ \textrm{OIII} \right]$ 4959 & D & 0.37 & 0.8 & 536.22 & 1.77 \\ 
$\left[ \textrm{OIII} \right]$ 4959 & E & -39.31 & 1.54 & 369.21 & 3.41 \\ 
$\left[ \textrm{OIII} \right]$ 4959 & X & -- & -- & -- & -- \\ 
$\left[ \textrm{OIII} \right]$ 5007 & A & 584.13 & 1.82 & 718.46 & 4.01 \\ 
$\left[ \textrm{OIII} \right]$ 5007 & B & 517.47 & 0.65 & 844.69 & 1.31 \\ 
$\left[ \textrm{OIII} \right]$ 5007 & C & 103.2 & 0.85 & 787.07 & 1.86 \\ 
$\left[ \textrm{OIII} \right]$ 5007 & D & -0.54 & 0.36 & 608.59 & 0.77 \\ 
$\left[ \textrm{OIII} \right]$ 5007 & E & -39.58 & 0.66 & 397.16 & 1.4 \\ 
$\left[ \textrm{OIII} \right]$ 5007 & X & -- & -- & -- & -- \\ 
GBand & A & -- & -- & -- & -- \\ 
GBand & B & 324.67 & 52.79 & 774.65 & 125.3 \\ 
GBand & C & -- & -- & -- & -- \\ 
GBand & D & 389.5 & 135.74 & 644.38 & 320.19 \\ 
GBand & E & 340.49 & 118.89 & 720.71 & 280.34 \\ 
GBand & X & -- & -- & -- & -- \\ 
MgIb & A & 232.57 & 56.8 & 477.78 & 118.44 \\ 
MgIb & B & 302.44 & 36.23 & 530.51 & 75.79 \\ 
MgIb & C & 99.39 & 43.86 & 588.35 & 93.74 \\ 
MgIb & D & 10.96 & 26.44 & 409.69 & 55.48 \\ 
MgIb & E & -2.53 & 7.44 & 14.73 & 12.84 \\ 
MgIb & X & 44.01 & 11.53 & 167.59 & 24.89 \\ 
\hline
\end{tabular}
\label{T:lines_1502b}
\end{table}

\begin{table}
\caption{Table \ref{T:lines_1502b} continued.}
\begin{tabular}{ccrrrr}
\hline
Line & Slice & v & $\Delta$ v & $FWHM$ & $\Delta FWHM$ \\
         &          &  [km s$^{-1}$] &  [km s$^{-1}$]  & [km s$^{-1}$] &  [km s$^{-1}$]  \\
\hline
GBand & A & -- & -- & -- & -- \\ 
GBand & B & 324.67 & 52.79 & 774.65 & 125.3 \\ 
GBand & C & -- & -- & -- & -- \\ 
GBand & D & 389.5 & 135.74 & 644.38 & 320.19 \\ 
GBand & E & 340.49 & 118.89 & 720.71 & 280.34 \\ 
GBand & X & -- & -- & -- & -- \\ 
MgIb & A & 232.57 & 56.8 & 477.78 & 118.44 \\ 
MgIb & B & 302.44 & 36.23 & 530.51 & 75.79 \\ 
MgIb & C & 99.39 & 43.86 & 588.35 & 93.74 \\ 
MgIb & D & 10.96 & 26.44 & 409.69 & 55.48 \\ 
MgIb & E & -2.53 & 7.44 & 14.73 & 12.84 \\ 
MgIb & X & 44.01 & 11.53 & 167.59 & 24.89 \\ 
\hline
\end{tabular}
\label{T:lines_1502c}
\end{table}

\begin{table}
\caption{Spatially resolved kinematics of lines for SDSS1715+6008. Continued in Table \ref{T:lines_1715b}. The lines are fitted in five spatial bins across the slit, as well as one additional bin for objects in which either tidal tails or a nearby neighbour falls on the slits. The location of those spatial bins are indicated in Fig. \ref{F:slitmaps} and are identical to those shown in Figures \ref{F:lines_0952} and \ref{F:spec_0952}. All velocities are given with respect to the nominal object redshift listed in Table \ref{T:objects}. Negative velocities indicate a blueshift and positive velocities a redshift.}
\begin{tabular}{ccrrrr}
\hline
Line & Slice & v & $\Delta$ v & FWHM & $\Delta$ FWHM\\
    &          &  [km s$^{-1}$] &  [km s$^{-1}$]  & [km s$^{-1}$] &  [km s$^{-1}$]  \\
\hline
Gas & A & -- & -- & -- & -- \\ 
Stars & A & -- & -- & -- & -- \\ 
Gas & B & -39.81 & 12.58 & 547.61 & 31.71 \\ 
Stars & B & 179.05 & 13.1 & 466.49 & 35.37 \\ 
Gas & C & -23.31 & 9.98 & 587.84 & 24.94 \\ 
Stars & C & 130.12 & 10.08 & 546.96 & 31.26 \\ 
Gas & D & 62.52 & 13.02 & 567.47 & 33.3 \\ 
Stars & D & 96.5 & 10.69 & 492.26 & 32.58 \\ 
Gas & E & -- & -- & -- & -- \\ 
Stars & E & -- & -- & -- & -- \\ 
$H_{\beta}$ & A & -109.42 & 13.21 & 187.14 & 30.81 \\ 
$H_{\beta}$ & B & 7.27 & 4.62 & 557.65 & 10.65 \\ 
$H_{\beta}$ & C & 94.43 & 4.25 & 663.14 & 9.44 \\ 
$H_{\beta}$ & D & 190.07 & 3.19 & 493.04 & 7.32 \\ 
$H_{\beta}$ & E & -- & -- & -- & -- \\ 
\hline
\end{tabular}
\label{T:lines_1715}
\end{table}

\begin{table}
\caption{Table \ref{T:lines_1715} continued.}
\begin{tabular}{ccrrrr}
\hline
Line & Slice & v & $\Delta$ v & FWHM & $\Delta$ FWHM\\
    &          &  [km s$^{-1}$] &  [km s$^{-1}$]  & [km s$^{-1}$] &  [km s$^{-1}$]  \\
\hline
$\left[ \textrm{OIII} \right]$ 4959 & A & -91.6 & 4.7 & 223.65 & 10.89 \\ 
$\left[ \textrm{OIII} \right]$ 4959 & B & -21.28 & 1.19 & 425.72 & 2.64 \\ 
$\left[ \textrm{OIII} \right]$ 4959 & C & 80.2 & 1.52 & 621.83 & 3.18 \\ 
$\left[ \textrm{OIII} \right]$ 4959 & D & 170.54 & 1.3 & 515.54 & 2.85 \\ 
$\left[ \textrm{OIII} \right]$ 4959 & E & 241.98 & 1.59 & 191.48 & 3.56 \\ 
$\left[ \textrm{OIII} \right]$ 5007 & A & -91.74 & 1.75 & 227.37 & 3.96 \\ 
$\left[ \textrm{OIII} \right]$ 5007 & B & -6.16 & 0.61 & 472.16 & 1.3 \\ 
$\left[ \textrm{OIII} \right]$ 5007 & C & 80.52 & 0.72 & 643.78 & 1.44 \\ 
$\left[ \textrm{OIII} \right]$ 5007 & D & 153.24 & 0.66 & 560.35 & 1.37 \\ 
$\left[ \textrm{OIII} \right]$ 5007 & E & 240.81 & 0.64 & 186.32 & 1.37 \\ 
$\left[ \textrm{OI} \right]$ 6300 & A & -- & -- & -- & -- \\ 
$\left[ \textrm{OI} \right]$ 6300 & B & 89.18 & 5.12 & 759.43 & 11.61 \\ 
$\left[ \textrm{OI} \right]$ 6300 & C & 126.89 & 3.61 & 804.3 & 8.08 \\ 
$\left[ \textrm{OI} \right]$ 6300 & D & 145.77 & 4.41 & 724.53 & 10.05 \\ 
$\left[ \textrm{OI} \right]$ 6300 & E & -- & -- & -- & -- \\ 
$\left[ \textrm{OI} \right]$ 6364 & A & 445.07 & 5.13 & 127.19 & 11.81 \\ 
$\left[ \textrm{OI} \right]$ 6364 & B & 117.43 & 18.24 & 857.73 & 42.45 \\ 
$\left[ \textrm{OI} \right]$ 6364 & C & 142.95 & 11.79 & 787.66 & 27.27 \\ 
$\left[ \textrm{OI} \right]$ 6364 & D & 159.31 & 14.07 & 744.36 & 33.0 \\ 
$\left[ \textrm{OI} \right]$ 6364 & E & -- & -- & -- & -- \\ 
\hline
\end{tabular}
\label{T:lines_1715b}
\end{table}

\section{Notes on Individual Sources}
\label{S:individual}

In this section detailed notes on individual sources not discussed in Section \ref{S:results} are presented.

\subsection{SDSS0952+2552}
\label{S:0952}

SDSS0952+2552 has a redshift of z=0.339, the host galaxy redshift from absorption is z=0.33892 \citep{smith_search_2010}. The velocity separation between the two $\left[ \textrm{OIII} \right]$ peaks is 475 km s$^{-1}$ \citep{smith_search_2010}. The object appears inconspicuous in the SDSS images and slightly kidney-shaped in the deep GTC $r$-band data.

AO NIR imaging shows two cores separated approximately in a N-S direction by 4.9 kpc \citep{fu_mergers_2011}, one of the NIR nuclei is associated with a Type 1 spectrum, the other with a Type 2 \citep{fu_nature_2012}. The AGN component and host galaxy are separated by 0.72\arcsec / 3.5 kpc, also in a N-S direction. In the residual, there is a weak compact source towards the south of the main point source that coincides with the second fainter AGN. This weak compact source coincides with slightly redder colors in $r-z$ and is detected in $u$. The orientation of the host galaxy is at an angle of 36$^\circ$ east of north.

There is a remarkably bright, very elongated galaxy towards the NW of the main galaxy, with a separation of about 40 kpc. No clear connecting bridge is observed between the two, but clear narrow line emission in H$\beta$ and $\left[ \textrm{OIII} \right]$ 4959/5007 are detected (Figures \ref{F:lines_0952}, \ref{F:spec_0952}). 

\subsection{SDSS1151+4711}
\label{S:1151}

SDSS1151+4711 has a redshift of z=0.31794 (using the red $\left[ \textrm{OIII} \right]$ peak), no host galaxy redshift can be determined from SDSS spectra \citep{smith_search_2010}. The velocity separation between the two $\left[ \textrm{OIII} \right]$ peaks is 490 km s$^{-1}$ \citep{smith_search_2010}. The host  shows very clear merger features with two strong tidal tails. The tidal tail towards the north spans almost 50 kpc, the tidal tail towards the east is considerably shorter, but still spans about 20 kpc.

NIR AO imaging shows the two AGN separated by 4.7 kpc at a position angle of about 45$^\circ$ west of north \citep{fu_nature_2012}. IFU data shows one Type 1 and one Type 2 AGN \citep{fu_nature_2012}. The galaxy is oriented at 38$^\circ$ east of north. The point source and galaxy are separated by 0.5\arcsec (2.3 kpc). There is a small clump about 10 kpc west of the core. There is an additional galaxy in the field towards the NW. The brighter point source towards the SW is listed as a star in SDSS and shows proper motion. 

The companion galaxy towards the NW of the main object is included in the slit but lacks clear line detection in either absorption or emission. The two $\left[ \textrm{OIII} \right]$ components are not clearly spatially separated.

\subsection{SDSS1502+1115}
\label{S:1502}

SDSS1502+1115 has a redshift of z=0.39326 (using the red $\left[ \textrm{OIII} \right]$ peak). No accurate host galaxy redshift from stellar absorption lines can be determined from SDSS spectra \citep{smith_search_2010}. The velocity separation between the two $\left[ \textrm{OIII} \right]$ peaks is 350 km s$^{-1}$ \citep{smith_search_2010}. The source appears clearly extended and has a double nucleus in the SDSS data. SDSS1502+1115 was observed in the radio \citep{fu_kiloparsec-scale_2011} and has been confirmed as a binary AGN.

AO NIR imaging shows two components separated by 1.39\arcsec/7.4 kpc \citep{fu_mergers_2011}. This is the object with the largest separation in our sample. IFU data shows one Type 1 and one Type 2 AGN \citep{fu_nature_2012}. The galaxy components are separated by 1.7\arcsec/9 kpc. No tidal tails or shells are detected. 

SDSS1502+1115 is the object in our sample with the fewest stars in the field usable for photometric calibration as well as alignment between the bands, therefore, the exact location of the slit shown in Fig. \ref{F:slitmaps} is somewhat unsure in the N-S direction.

Several compact clumps are detected around the host galaxy. A very compact source about 30 kpc east of the core was included in the long slit spectroscopy and is at the same redshift as the AGN, it shows clear absorption in H$\beta$, H$\delta$, H$\gamma$, G Band and MgIb.

\subsection{SDSS1715+6008}
\label{S:1715}

SDSS1715+6008 has a host galaxy redshift of z=0.15648 \citep{smith_search_2010}. The velocity separation between the two $\left[ \textrm{OIII} \right]$ peaks is 720 km s$^{-1}$ \citep{smith_search_2010}. X-ray observations reveal a double source, confirming that the source likely hosts a binary AGN \citep{comerford_chandra_2011}. The two AGN are separated by 0.68\arcsec/1.85 kpc in a south-east direction \citep{comerford_chandra_2011}. The host galaxy appears undisturbed  even in the deep GTC data.

All bands require a point source to achieve a good fit, though the AGN is considerably fainter than the host in all bands. The host galaxy is best fit by a disk bulge mixture in $r$ and $z$. The $u$-band data reveals only a marginal detection of the host galaxy. Only the $r$ band residual reveals a very weak disturbed morphology in the central regions. It should be noted that $\left[ \textrm{OIII} \right]$ is located in the center of the $r$-band, therefore strong $\left[ \textrm{OIII} \right]$ emission will result in structures in the $r$ band morphology.

The spectrum of SDSS1715+6008 covers a rest-frame wavelength range of about 4866--6516$\AA$. H$\alpha$ and [NII]$\lambda$6583$\AA$ are in the spectral range but are too heavily affected by telluric absorption to be of use for further analysis.

\end{document}